\documentclass[12pt]{article}

\usepackage{amsmath}         
\usepackage{amssymb}         
\usepackage{amsfonts}        
\usepackage{graphicx}        

\usepackage{cite}          
\usepackage{youngtab}	   
\usepackage{mathrsfs}      

\usepackage[margin=2cm]{geometry} 

\graphicspath{{figures/}}	        
\numberwithin{equation}{section}    

\renewcommand{\tilde}{\widetilde}

\renewcommand{\vec}[1]{\mathbf{#1}}

\newcommand{\f}{\Box}
\newcommand{\af}{\overline{\Box}}
\newcommand{\anti}{\drawsquare{4}{0.6}\hskip-4.6pt%
	\raisebox{4pt}{\drawsquare{4}{0.6}}}
\newcommand{\antin}{\tiny \Yvcentermath1 \yng(1,1)}
\newcommand{\antim}{{\fontsize{2}{3}\selectfont \Yvcentermath1 \yng(1,1)}}
\newcommand{\aanti}{{\fontsize{2}{3}\selectfont \Yvcentermath1 \overline{\yng(1,1)}}}
\newcommand{\I}{1}
\newcommand{\beq}{\begin{equation}}
\newcommand{\eeq}{\end{equation}}
\newcommand{\bea}{\begin{eqnarray}}
\newcommand{\eea}{\end{eqnarray}}

\newcommand{\drawsquare}[2]{\hbox{%
\rule{#2pt}{#1pt}\hskip-#2pt
\rule{#1pt}{#2pt}\hskip-#1pt
\rule[#1pt]{#1pt}{#2pt}}\rule[#1pt]{#2pt}{#2pt}\hskip-#2pt
\rule{#2pt}{#1pt}}

\newenvironment{institutions}[1][2em]
  {\begin{list}{}{\setlength\leftmargin{#1}\setlength\rightmargin{#1}}\item[]}
  {\end{list}}

\usepackage[
	colorlinks=true,
	citecolor=black,
	linkcolor=black,
	urlcolor=blue,
	hypertexnames=false]{hyperref}

\begin{document}

\thispagestyle{empty}
\begin{center}

	{	
		\LARGE \bf 
		Dynamics of 3D SUSY Gauge Theories\\ \vspace*{0.2cm}
        with Antisymmetric Matter
	}

	\vskip .7cm
	
	\renewcommand*{\thefootnote}{\fnsymbol{footnote}}

	{	
		\bf
		Csaba Cs\'aki$^{a}$, Mario Martone$^{a}$, Yuri Shirman$^{b}$,\\
		\vspace{.2cm}
		Philip Tanedo$^{b}$, and {John Terning}$^{c}$
	}  

	\vspace{.2cm}

\begin{institutions}[2.25cm]
\footnotesize

    
	$^{a}$ {\it Dept.\ of Physics, \textsc{lepp}, Cornell University, Ithaca, NY 14853}  
	\\
	$^{b}$ {\it Dept.\ of Physics \& Astronomy, University of California, Irvine, CA 92697} 
	\\
	$^{c}$ {\it Dept.\ of Physics, University of California, Davis, CA 95616} 
\end{institutions}	
\end{center}

\vspace{-.5em}
\begin{center}{ \footnotesize\tt csaki@cornell.edu, mcm293@cornell.edu, yshirman@uci.edu, \\ flip.tanedo@uci.edu, jterning@gmail.com} 
\end{center}


\begin{abstract}
\noindent 
We investigate the IR dynamics of ${\cal N}=2$ SUSY gauge theories in 3D with antisymmetric matter. The presence of the antisymmetric fields leads to further splitting of the Coulomb branch. Counting zero modes in the instanton background suggests that more than a single direction along the Coulomb branch may remain unlifted. We examine the case of $SU(4)$ with one or two antisymmetric fields and various flavors in detail. Using the results for the corresponding 4D theories, we find the IR dynamics of the 3D cases via compactification and a real mass deformation. We find that for the s-confining case with two antisymmetric fields, a second unlifted Coulomb branch direction indeed appears in the low-energy dynamics. We present several non-trivial consistency checks to establish the validity of these results. We also comment on the expected structure of general s-confining theories in 3D, which might involve several unlifted Coulomb branch directions. 
\end{abstract}


\section{Introduction}

The understanding of the dynamics of SUSY gauge theories was revolutionized in the mid-1990s due to the seminal work of Seiberg and Intriligator on $\mathcal N=1$ theories~\cite{Seberg:1994conf,Seiberg:1994pq, Intriligator:1995au}, and Seiberg and Witten for ${\cal N}=2$ theories~\cite{Seiberg:1994sw1,Seiberg:1994sw2}.
These works show how the interplay of holomorphy, global symmetries, instantons, anomalies, and monopoles can determine the IR behavior of a large class of theories. 
Depending on the amount of matter, these theories manifest dynamical effects such as gaugino condensation, instanton generated superpotentials, confinement with or without chiral symmetry breaking, IR-free composite gauge groups, interacting non-Abelian quantum fixed points, monopole condensation as a dual of confinement, etc. 

While the initial work focused on 4D gauge theories, Aharony, Seiberg, Intriligator and many others extended some of these results to 3D ${\cal N}=2$ theories, where many of the corresponding phenomena have also been observed~\cite{
Seiberg:1996comp,
Intriligator:1996mi,
deBoer:1997kr,
Aharony:1997bx,
Karch:1997ux,
Aharony:1997ad,
Kapustin:1999ha,
Strassler:1999conf,
Dorey:2000mirror,
Tong:2000CS,
Giveon:2008gk,
Niarchos:2008CS,
Kapustin:2011gh,
Dimofte:2011,
Benini:2011mf,
Aharony:2011ci,
Hwang:2011dual,
Kapustin:2011vz,
Kim:2013}.  
There are many reasons why one would investigate lower dimensional theories. 3D can be a simpler laboratory for more complicated dynamics in 4D, for example, as a tool for understanding possible mechanisms for confinement \cite{Polyakov:1976fu, Feynman:1981ss}.
In fact, a new mechanism for confinement from bion condensation was proposed~\cite{Unsal:2007jx,Unsal:2007vu} and further studied in~\cite{Argyres:2012ka,Poppitz:2011wy,Poppitz:2009uq,Shifman:2009tp}.
Since one may generate 3D theories from the compactification of 4D theories, one can expect that some of their behavior is reflective of 4D properties. On the other hand, 3D quantum field theories also have novel elements not found in 4D such as instanton-monopoles, Chern-Simons terms, topological phases, and real masses \cite{Aharony:1997bx,deBoer:1997kr}.
These may have applications to condensed matter systems, see for example a recent application of 3D mirror symmetry to Fermi surfaces \cite{Hook:2014dfa}.

Over the past year, compelling derivations of many of the results obtained in~\cite{
Karch:1997ux,
Aharony:1997ad,
Aharony:1997bx,
deBoer:1997kr,
Giveon:2008gk,
Benini:2011mf,
Aharony:2011ci
} 
have emerged \cite{Aharony:2013dha,Aharony:2013hda,Park:2013}. A careful sequence of compactification $\mathbb{R}^4 \to \mathbb{R}^3\times S^1$ together with a real mass deformation yields purely three dimensional electric theories and allows one to determine the magnetic duals in a controlled way\footnote{See Appendix~\ref{app:dim:reduction} for a self-contained introduction to dimensional reduction of 4D dualities. Also see \cite{Niarchos:2012sd}.} \cite{Aharony:2013dha}.
The main purpose of this paper is to take the first step to extend these results to theories with more general matter. In particular, we focus on the case with antisymmetric tensors. A careful counting of the fermionic zero modes in instanton backgrounds leads us to conclude that the dimension of the unlifted Coulomb branch can be larger once representations other than fundamentals are included. The reason for this is that the charges of the moduli parameterizing the Coulomb branch are modified in the presence of antisymmetric matter leading to fewer directions lifted by instanton effects.
In the specific example of an $SU(4)$ gauge group we identify two unlifted directions: the standard $Y$ direction found in the presence of fundamental matter fields and a new $\tilde{Y}$ direction arising from the additional splitting of the Coulomb branch due to massless components of the antisymmetric matter fields---this direction is related to the unlifted Coulomb branch modulus of $SO(6)$ theories with vectors. Using this insight we apply the program of~\cite{Aharony:2013dha} 
to the simplest examples of 4D models with antisymmetric matter and known 4D dynamics: 
the s-confining~\cite{Csaki:1996zb,Csaki:1996sm,Csaki:1998th} $SU(4)$ gauge theories with either two antisymmetrics and three flavors or one antisymmetric and four flavors. We find that for the case of two antisymmetrics, the corresponding s-confining 3D theory with two flavors indeed requires the $\tilde{Y}$ operator in the description of the IR dynamics. We perform several consistency checks, including the matching of quantum numbers, reproducing the quantum modified constraints expected for theories on a circle, and connecting it to theories with fewer flavors, establishing a consistent, intricate web of IR dynamics of several $SU(4)$ theories with antisymmetric matter and flavors. Our experience with these theories leads us to speculate on the classification of the Coulomb branch structure for general s-confining theories in 3D.

The paper is organized as follows. We review the dynamics of the Coulomb branch for theories with fundamental matter fields in Section 2. Section 3 contains a discussion of the structure of the Coulomb branch in the presence of antisymmetric matter. We apply the program of~\cite{Aharony:2013dha} to the s-confining SU(4) theories with antisymmetric matter in Section 4 to find the correct low-energy dynamics of those models, and perform various consistency checks. In Section 5 we summarize our expectations for the general behavior of 3D s-confining theories, and finally conclude in Section 6. The appendices introduce 3D ${\cal N}=2$ SUSY gauge theories (Appendices A and B), zero mode counting and the Callias index theorem (Appendix~C), and summarize the 3D dynamics of SUSY QCD theories with various number of flavors (Appendix~D).

\section{Review of Coulomb branch dynamics}\label{sec:CB}

We summarize the behavior of the Coulomb branch in the presence of matter representations; see \cite{Aharony:1997bx,deBoer:1997kr} for thorough reviews. As detailed in Appendix~\ref{app:notation}, the 3D vector multiplet has a scalar component, $\sigma$, which has no potential in the absence of additional matter fields. Thus $\sigma$ is a flat direction and can be diagonalized by a gauge transformation.  The general scalar VEV can be written as $\langle\sigma\rangle=\sigma^iH_i$, where $H_i$ are elements of the Cartan subalgebra. This VEV breaks the gauge group to its maximal Abelian subgroup, e.g.~$SU(N)\to U(1)^{N-1}$. 
In order to remove the leftover gauge redundancy, one must impose additional constraints. The most common choice is to assume
\beq
\vec{\alpha}_k\cdot  \vec{\sigma} \geq 0 \qquad\qquad \left(k=1,...,r_G\right),
\eeq
where the $\vec{\alpha}_k$ are the simple roots and $r_G$ is the rank of the group. These conditions restrict $\sigma$ to a Weyl chamber.
Unless otherwise stated, we focus on $SU(N)$, which is broken to $U(1)^{N-1}$ in the bulk Coulomb branch. In this case, the Weyl chamber is described by the subspace
\beq\label{higgs:vacuum}
\sigma={\rm diag}(\sigma_1,\sigma_2,...,\sigma_{N}),\qquad\qquad \sigma_1\geq\sigma_2\geq \cdots \geq\sigma_{N}~.
\eeq 
The eigenvalues $\sigma_i$ are not completely independent since $\sigma$ is traceless,  $\sigma_1+\cdots+\sigma_N=0$. 

In the absence of matter, the Coulomb branch is a cylinder, $\mathbb{R}\times S^1$, described by $(N-1)$ moduli,
\beq\label{inst}
Y_i\sim \exp{\left(\vec{\Phi}\cdot\boldsymbol{\alpha}_i/g_3^2\right)},
\eeq
where $g_3$ is the 3D gauge coupling and $\boldsymbol{\alpha}_i$ are the simple roots.
$\vec{\Phi}$ is the chiral superfield whose lowest component contains the dual photon and the scalar $\sigma$. It is dual to the linear superfield that encodes the gauge degrees of freedom, as reviewed in Appendix~\ref{app:notation}.
In the remainder of this section we show how this classical picture is modified both by the presence of matter fields and non-perturbative corrections. 
%

\subsection{Coulomb branch pinch from matter}

Matter fields, $Q_i$ and $\bar Q_i$, couple to the lowest component of the vector superfield, $\sigma$, according to
\beq
\sim 
\left|
\left(\sigma^a T^a_Q\right)^\alpha_{\phantom{\alpha}\beta} 
Q^\beta_{\phantom{\beta}i}
\right|^2,
\eeq
where the $T^a$ are group generators in the representation of the $Q$ fields, $\alpha$ and $\beta$ are gauge indices, and $i$ is a flavor index. 
On the Coulomb branch, the scalar VEV is an effective real mass for the matter fields,
\begin{align}
\label{higgs:effmass}
\left|
\langle\sigma^a T^a\rangle^\alpha_{\phantom{\alpha}\beta} 
Q^\beta_{\phantom\beta i} 
\right|^2
& \sim 
\left|
\langle\vec\sigma\rangle \cdot \vec\nu_{\alpha} \,
Q^\alpha_{\phantom\beta i}\right|^2
\qquad\quad
\text{\footnotesize (no sum over $\alpha$)}
\end{align}
where 
$\vec{\nu}_\alpha$ is the weight of the $\alpha^{\text{th}}$ vector in the $Q$
representation. Due to this real mass, the Higgs branch only intersects the Coulomb branch when $\vec{\sigma}\cdot\vec{\nu}_\alpha=0$.

\begin{figure}
\begin{center}
\includegraphics[width=.57 \textwidth]{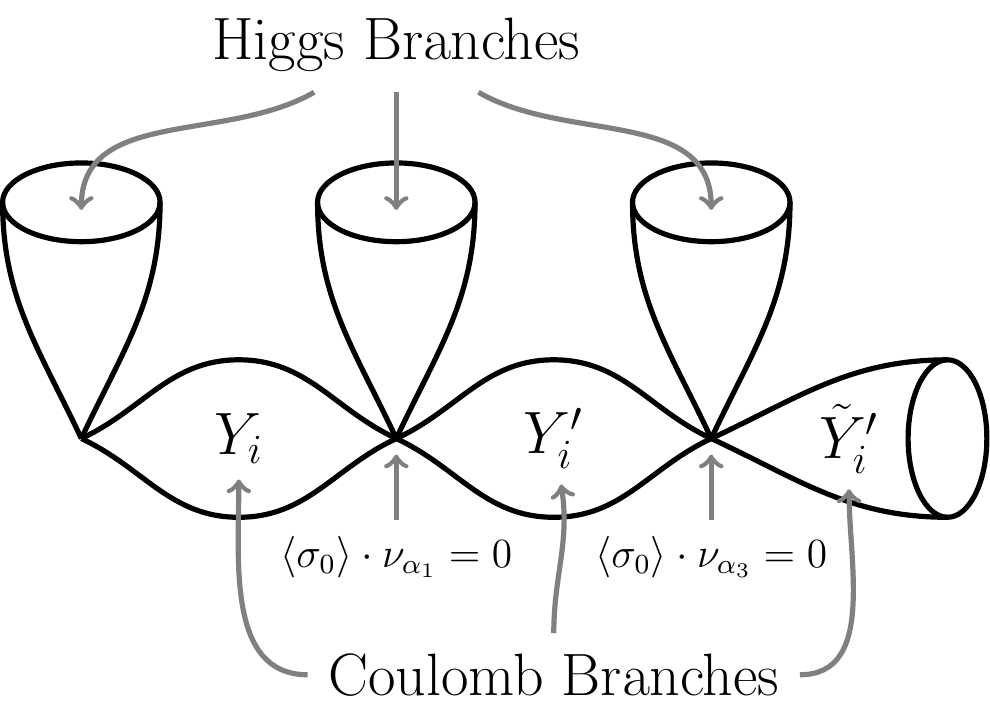}
\end{center}
\caption{In presence of matter, the Coulomb branches pinch off where they meet Higgs branches. At these points, $\vec{\sigma}\cdot\vec{\nu}_\alpha=0$ and there is massless matter.}
\label{fig:pinches}
\end{figure}

At these points, the low energy theory is governed by a particular $U(1)\subset SU(N)$ with $F$ massless flavors.
When including quantum effects, the Coulomb branch `pinches' and splits into two distinct regions where it intersects the Higgs branch, see Fig.~\ref{fig:pinches}. The two regions are $\sigma>0$ ($\mathbb R^+\times S^1$) and $\sigma<0$ ($\mathbb R^-\times S^1$) \cite{Aharony:1997bx, Intriligator:2013lca} and are described by two independent operators $V^{\pm}$ \cite{Aharony:1997bx}. 
We expect similar dynamics to take place in the $SU(N)$ case at each intersection of the Coulomb branch with a Higgs branch so that the Coulomb branch splits along this sub-locus with different variables describing either side of each boundary. The operators $V^\pm$ are identified with a particular combination of the chiral operators $Y_i$ in (\ref{inst}) on the two sides of each singularity. 


\subsection{Instanton/monopoles, $Y_i$ charges, global coordinates}

The moduli $Y_i$ are not viable global coordinates for the Coulomb branch. This is a consequence of the facts that the $Y_i$ are charged under global $U(1)$ symmetries and that these charges jump at the singularities. To see this, one may compute the charges of the moduli by counting the fermion zero modes in 3D instanton backgrounds.
To that end, note that 3D instanton solutions can be obtained by starting with 4D monopoles and compactifying the Euclidean time coordinate.
Thus 3D instantons are simply 4D 't~Hooft-Polyakov monopoles with the scalar component of the vector superfield $\sigma$ corresponding to the adjoint Higgs scalar\footnote{The same result can be seen from homotopy theory. In 3D, the classification of instantons is based on $\pi_2(G)$. Since the scalar component of the vector superfield triggers the breaking $SU(N)\to U(1)^{N-1}$, we expect to find $(N-1)$ instantons $\pi_2(G/U(1)^{N-1})\cong \pi_1(U(1)^{N-1})\cong \mathbb{Z}^{N-1}, $ which makes the connection to 4D monopoles evident.}.
To emphasize this 4D intuition, we refer to these 3D instantons as instanton/monopole solutions. 
See Appendix~\ref{app:callias} or the original literature \cite{Weinberg:1980mo,Weinberg:1982mo} for details. 
We thus expect a 3D instanton solution associated with each $SU(2)$ subgroup. In particular, $SU(N)$ has $(N-1)$ independent solutions, one for each $SU(2)$ subgroup. These are labeled by topological indices $(n_1,n_2,...,n_{N-1})$ which are the charges of the independent 4D BPS monopoles.

The operator $Y_i$ in (\ref{inst}) is associated with the $i^\text{th}$ fundamental instanton for which $n_k=\delta_{ik}$. 
If a fermion charged under a global $U(1)$ has zero modes in this particular instanton/monopole background, $Y_i$ also acquires charges under that $U(1)$. Because the zero mode counting changes in different Coulomb branch regions, the charges of the $Y_k$ under the global $U(1)$s change as well. The holomorphic structure is different in each region and there is no reason to assume that a given $Y_k$ is continuous across a singularity. We therefore conclude that if, for a given index $i$, the number of zero modes in an 
$n_k=\delta_{ik}$
instanton/monopole background jumps while crossing from one region to another, the operator $Y_i$ is discontinuous at the pinch and we need two independent operators, $Y_i$ and $Y_i'$, to describe the Coulomb branch on each side of the pinch.

There exists, however, a set of globally defined operators that parameterize the Coulomb branch. One such operator is
\begin{align}
Y=\prod_{i=1}^{N-1} Y_i.
\label{eq:Y:prod:Yi}
\end{align}
The continuity of $Y$ introduces an extra constraint among the $Y_k$ at each split. The number of globally defined moduli needed to parameterize the Coulomb branch is determined both by the number of singularities and the number of moduli whose charges jump at each of these singularities\footnote{
This statement requires some care. Even in absence of real masses, $Y$ might not be globally defined in $SU(N)$ theories with $N>4$ and with generic matter representations. Throughout the core of this paper, $Y$ is continuous and the statement above is true for vanishing real masses. We leave the study of $N>4$ for future work~\cite{Csaki:2014new}.
}.

\subsection{Monopole operators}\label{sec:monopole}

3D gauge theories with $U(1)$ factors have a topological global symmetry, $U(1)_{\mathcal{J}}$, whose associated conserved current is $\mathcal{J}_i\equiv*F_i=\epsilon_{ijk}F^{jk}$. See \cite{Aharony:1997bx,Intriligator:2013lca} or Appendix~\ref{app:notation} for details. The associated charge is
\beq
q_\mathcal{J}=\int d^2x\; j_{\mathcal{J}}^0=\int d^2x\; F_{12}.
\eeq
A non-vanishing charge $q_J$ therefore introduces a magnetic flux. Operators that introduce a magnetic flux through any sphere ${S}^2$ surrounding a spacetime point $x$, are called monopole operators 
\cite{
Kapustin:2005py,
Borokhov:2002cg,
Borokhov:2002ib,
Aharony:2013kma,
Intriligator:2013lca,
Aharony:2013dha,
Intriligator:2014mon}. 
Subject to some subtleties, these monopole operators act on the vacuum to create BPS vortices---topologically charged states localized at a point. 

As explained in Appendix~\ref{app:notation}, the photon of any unbroken $U(1)$ is dual to a periodic scalar, $\gamma$. The operator $\mathcal{O}_1(x)\sim \exp{[i\gamma(x)]}$ creates a magnetic flux of charge $q_J=1$.
The $SU(N)$ gauge group, on the other hand, is simple and does not contain any $U(1)$ factors. Monopole operators can only appear on the Coulomb branch where $SU(N)$ is broken to $U(1)$ subgroups. 
In particular, we show below that the UV monopole operators are those that parameterize the un-lifted Coulomb branch directions.
%

\subsection{Non-perturbative corrections}
\label{sec:nonpert:coulomb}

In 3D $\mathcal{N}=2$ super Yang Mills theories, instanton  configurations generate non-perturbative corrections to the superpotential. The $i^\text{th}$ fundamental instanton configuration, 
$n_k=\delta_{ik}$, 
contributes \cite{Affleck:1982as} 
\beq
W_\text{inst}\sim Y_i^{-1}\,.
\label{eq:AHW}
\eeq
The full superpotential is given by the sum over $(N-1)$ instanton contributions. In the absence of matter fields, the Coulomb branch is thus completely lifted. 
A simple way to understand the origin of this superpotential is to count the fermionic zero modes in the presence of the instanton~\cite{deBoer:1997kr}. In a pure super Yang Mills theory, each instanton supports two gaugino zero modes and implies an instanton-generated interaction that is quadratic in the gauginos. In the effective Lagrangian, such a term arises from the expansion of the superpotential (\ref{eq:AHW}) in superspace coordinates.

Light matter fields change the low energy dynamics.
Such fields may have zero modes in instanton configurations. 
These, in turn, prevent the corresponding superpotential term (\ref{eq:AHW}) from being generated and leave a part of the Coulomb branch unlifted~\cite{deBoer:1997kr,Aharony:1997bx,Affleck:1982as}
To illustrate this effect, we review the example of 3D SUSY QCD with $F$ flavors of fundamental matter:
\begin{align}
\begin{tabular}{l|c|ccc}
&$SU(N)$&$U(1)_A$&$U(1)_B$&$U(1)_R$\\
\hline
$Q$&$\square$&1&1&0\\
$\bar{Q}$&$\overline{\square}$&$1$&$-1$&0\\
\hline
$\lambda$&adj.&0&0&$1$\\
\end{tabular},
\end{align}
where we write the global U(1) symmetry charges of the matter superfields $Q,\bar{Q}$ as well as the gaugino $\lambda$. The flavor quantum numbers are not be relevant for the immediate discussion.

\begin{figure}
\begin{center}
\includegraphics[width=.67 \textwidth]{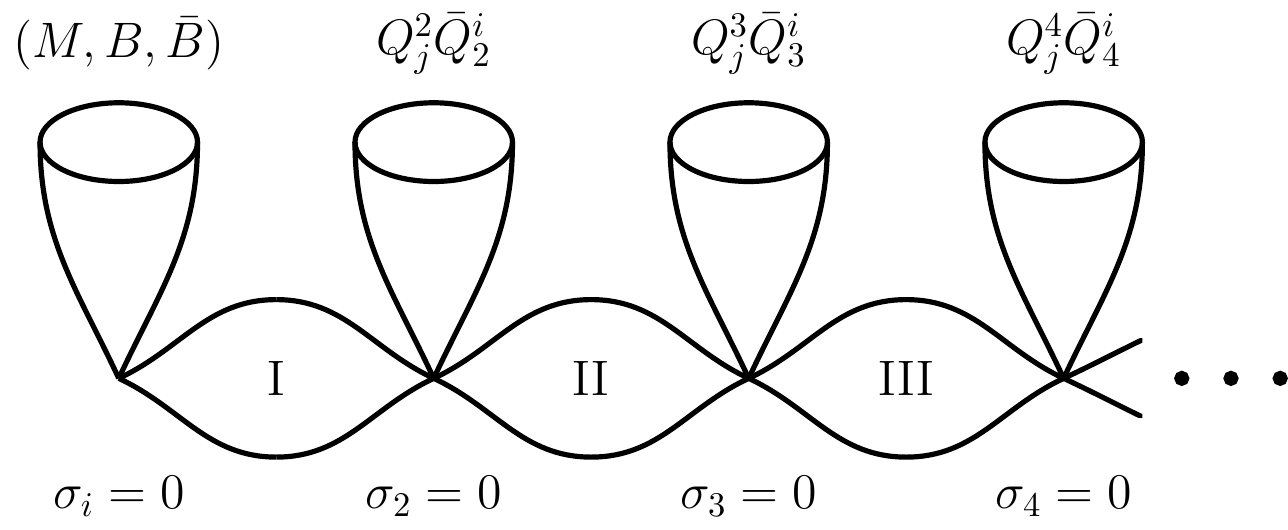}
\end{center}
\caption{In the theory with fundamentals, the Coulomb branch splits at each point where $\sigma_i=0$. The total Coulomb branch is composed of $(N-1)$ regions.}
\label{fig:fundamentals}
\end{figure}  

The Coulomb branch has singularities when a real mass for the matter fields vanishes. Since the real mass for the $i^\text{th}$ component of a fundamental is proportional to $\sigma_i$, the Coulomb branch splits into $(N-1)$ regions, as shown in Fig.~\ref{fig:fundamentals}. 
The $i^\text{th}$ region is defined by $$\sigma_1>...>\sigma_i>0>\sigma_{i+1}>...>\sigma_N.$$ 
We review the counting of fermion zero modes using the Callias index theorem in Appendix~\ref{app:callias}. 
Here we present a more intuitive description. Each instanton/monopole solution is associated with a particular $SU(2)$ subgroup of $SU(N)$, say the $i^\text{th}$ such subgroup. 
A quark decomposes into a doublet and $(N-2)$ singlets with respect to this subgroup.
On the Coulomb branch, this $SU(2)$ is broken to a $U(1)$ subgroup by the corresponding adjoint VEV which also sets the size of the instanton,
\beq\label{monopole:size}
\frac{1}{\rho^i_{\rm mon}}=\frac{1}{2} \big|\sigma_i-\sigma_{i+1}\big|\,.
\eeq 
The $\sigma$ VEVs induce a real mass for the $SU(2)$ doublet  $(Q^i,Q^{i+1})$ in $Q$. Since $Q$ is in the fundamental representation,
\beq\label{higgs:mass}
m^i_{\rm eff}=\frac{1}{2}\big|\sigma_i+\sigma_{i+1}\big|\,.
\eeq
If the effective doublet mass is larger than the inverse size of the instanton (\ref{monopole:size}), then the quarks do not have zero modes in the $i^\text{th}$ instanton background. On the other hand, if the inverse size of the instanton is larger than the effective real mass, each quark has $n_i$ zero modes, where $n_i$ is the instanton charge. For details see, for example, \cite{Shifman:2012book}.
(\ref{monopole:size}) and (\ref{higgs:mass}) imply that each fundamental contributes $n_i$ zero modes in the $i^\text{th}$ instanton background if and only if $\sigma_i>0>\sigma_{i+1}$.

This argument applies to any representation with an appropriate generalization of (\ref{higgs:mass}), as can be verified by the Callias index theorem. It is important to note that the number of zero modes depends on the fermion representation. In particular, matter fields in larger representations may simultaneously have zero modes under different instantons.
Carrying out a similar calculation for gauginos, one finds $2n_i$ zero modes under $i^\text{th}$ instanton/monopole independent of the size of the $\sigma_i$. Thus there are $2\sum_i n_i$ gaugino zero modes everywhere on the Coulomb branch.

In summary, the Coulomb branch of an $SU(N)$ theory with $F$ flavors splits into $(N-1)$ regions and in the $i^\text{th}$ region, where $\sigma_i>0>\sigma_{i+1}$, there are $2F n_i$ zero modes coming from the fundamental and anti-fundamental fermions, as well  $2\sum_i n_i$ gaugino zero modes. 
Thus the $U(1)$ charges for the $Y_k$ in the $i^\text{th}$ region are given by
\begin{align}
\sigma_i>0>\sigma_{i+1}>\sigma_{i+2}\qquad\to\qquad
\begin{tabular}{c|ccc}
&$U(1)_A$&$U(1)_B$&$U(1)_R$\\
\hline
$Y_{k\neq i}$&0&0&$-2$\\
$Y_i$&$2F$&0&$F-2$
\end{tabular}
\end{align}
We see that the presence of matter zero modes prevents the appearance of the $Y_i^{-1}$ term and the effective superpotential in the $i^\text{th}$ region is 
\begin{equation}
 W=\sum_{k\neq i}\frac{1}{Y_k}\,.
\end{equation}
Since this result applies to all values of $i$, the quantum numbers of some of the $Y$ operators change at each singularity. In particular, at a singularity defined by $\sigma_{i+1}=0$, two operators, $Y_i$ and $Y_{i+1}$ are discontinuous and one must introduce two new operators.  It is helpful to illustrate these results for an $SU(4)$ gauge group with $F$ flavors:
\begin{align}\label{sufour:example}
\begin{tabular}{r|c|c|c|c}
&${\rm Region}$&$\text{Zero\ Modes}$&${\rm Coulomb\ operators}$&$W$\\
\hline
$\vphantom{i^{i^{i^i}}}$
I &
$\sigma_1>0>\sigma_2>\sigma_3>\sigma_4$ &
$2F n_1$ &
$Y_1,\ Y_2,\ Y_3$ &
$Y^{-1}_2+Y_3^{-1}$
\\
$\vphantom{i^{i^{i^i}}}$
II &
$\sigma_1>\sigma_2>0>\sigma_3>\sigma_4$ & 
$2F n_2$ &
$\tilde{Y}_1,\ \tilde{Y}_2,\ Y_3$ & 
$\tilde{Y}_1^{-1}+Y_3^{-1}$
\\
$\vphantom{i^{i^{i^i}}}$
III & 
$\sigma_1>\sigma_2>\sigma_3>0>\sigma_4$ & 
$2F n_3$ &
$\tilde{Y}_1,\ Y'_2,\ \tilde{Y}_3$ &
$\tilde{Y}^{-1}_1+Y'^{-1}_2$
\end{tabular}
\end{align}
There are seven distinct $Y_k$ operators in total. Four of these operators appear in the dynamical superpotential and lift parts of the Coulomb branch. The continuity of a globally defined $Y$ operator imposes two constraints relating the operators. It follows that the unlifted Coulomb branch is parameterized by a single operator, $Y$~\cite{Aharony:1997bx}.

\section{The Coulomb branch with antisymmetric matter}

We now take the first steps towards extending the results of the previous section to a theory that incorporates more general matter representations. Depending on the choice of representations, the Coulomb branch may be described by more than a single chiral operator. 
We concentrate on a simple example: $SU(4)$ $\mathcal{N}=2$ SUSY gauge theory with matter in (anti-)fundamental ($Q,\bar Q$) and antisymmetric tensor representations ($A$). Many of our conclusions, however, are general. In this section we do not fix the exact number of antisymmetrics and fundamental flavors and instead focus on the general properties of the Coulomb branch in the presence of these two representations.

As explained in the previous section, the number of independent chiral operators which describe the Coulomb branch depends on the number of pinches that occur when a matter field becomes massless. As before, we restrict the scalar VEV to a single Weyl chamber:
\beq\label{Weyl}
\sigma_1\geq\sigma_2\geq\sigma_3\geq\sigma_4, \qquad\text{ and }\qquad  \sigma_1+\sigma_2+\sigma_3+\sigma_4=0\ .
\eeq
For simplicity we suppress the flavor indices since gauge interactions are flavor diagonal. The classical vacua satisfy
\begin{align}
\sigma_\alpha Q^\alpha &=0
&
-\sigma_\alpha \overline{Q}^\alpha&=0
&
\left(\sigma_\alpha + \sigma_\beta\right)A^{\alpha\beta} &= 0,
\end{align}
where $\alpha, \beta \in \{1, \cdots 4\}$ and there is no sum over repeated indices.
The Higgs branches pinch the Coulomb branches at $\sigma_\alpha=0$ and at $\sigma_\alpha+\sigma_\beta=0$. 
In light of (\ref{Weyl}), the non-trivial Coulomb branch singularities are $\sigma_2=0$, $\sigma_3=0$ and $\sigma_2+\sigma_3=0$. 
The last singularity is due to the presence of antisymmetric matter. The Coulomb branch splits into four distinct regions. The Callias index theorem tells us the number of fermion zero modes in each region and thus determines the nonperturbative contributions to the superpotential and the number of independent $Y_i$ operators. 
The results are summarized in the Section~\ref{sec:coulombcoords}, with a detailed derivation given in Appendix~\ref{app:callias}. The main result is that the each antisymmetric tensor provides two additional zero modes: if $\sigma_2+\sigma_3>0$ the zero modes are in the first fundamental instanton, while for  $\sigma_2+\sigma_3<0$ they are in the third\footnote{%
This calculation is performed in a generic background on the Coulomb branch. This is the reason why the result differs from the zero mode structure in~\cite{Shifman:2008conf}, which was calculated assuming a center-symmetric background instead.
}.

\subsection{Coulomb branch coordinates}
\label{sec:coulombcoords}
The intuitive picture for the zero mode counting works out the same way. For example, consider the zero modes under the first fundamental instanton/monopole, corresponding to the $SU(2)$ embedded into the upper left corner of $SU(4)$. The background VEV for the adjoint can be written as a sum of two contributions: 
\begin{align}
\text{diag}\left(\frac{\sigma_1-\sigma_2}{2}, -\frac{\sigma_1-\sigma_2}{2},0,0\right)
&&
\text{diag} \left(\frac{\sigma_1+\sigma_2}{2},\frac{\sigma_1+\sigma_2}{2}, \sigma_3, \sigma_4\right).
\end{align}
the first determines the instanton size while the second is invariant under the $SU(2)$ rotations and contributes to the effective real mass~\cite{Weinberg:1980mo,Weinberg:1982mo,Poppitz:2008it}.
The antisymmetric of $SU(4)$ decomposes into two doublets and two singlets.
The doublet masses are $\frac 12 (\sigma_1+\sigma_2) + \sigma_{3,4}$.
Comparing the doublet masses with the inverse instanton/monopole size (\ref{monopole:size}) we obtain the following results:
\begin{eqnarray}
\nonumber
&
SU(4) \text{ with }\ F_A\ \antim \, + \ F \left(\f+\af\right) &
\\ \label{CoulombSU(4)}
&
\begin{array}{r|c|c|c|c}
&
{\rm Region} &
{\rm Zero\ Modes} &
\begin{split}
    \mathrm{Coulomb} \\
    \mathrm{operators}
\end{split} &
W 
\\
\hline
\vphantom{i^{i^{i^i}}} 
\mathrm{I} &
\sigma_1>0>\sigma_2>\sigma_3>\sigma_4 &
\left(2F +2F_A\right)n_1 &
Y_1,\ Y_2,\ Y_3 &
Y^{-1}_2+Y_3^{-1}
\\
\vphantom{i^{i^{i^i}}}
\mathrm{II} &
\sigma_1>|\sigma_3|>\sigma_2>0>\sigma_3>\sigma_4 &
2F_An_1+ 2F n_2 &
\tilde{Y}_1,\ \tilde{Y}_2,\ Y_3 &
Y_3^{-1}
\\
\vphantom{i^{i^{i^i}}}
\mathrm{III} &
\sigma_1>\sigma_2>|\sigma_3|>0>\sigma_3>\sigma_4 &
2F n_2+2F_A n_3 &
Y'_1,\ \tilde{Y}_2,\ \tilde{Y}_3 &
Y'^{-1}_1
\\
\vphantom{i^{i^{i^i}}_{I_{I_I}}}
\mathrm{IV} &
\sigma_1>\sigma_2>\sigma_3>0>\sigma_4 &
\left(
2F+2F_A\right)n_3 &
Y'_1,\ Y'_2,\ Y'_3 &
Y'^{-1}_1+Y'^{-1}_2
\end{array} &
\end{eqnarray}

In regions II and III, the zero modes of the fundamentals and antisymmetrics are misaligned so that fewer directions are lifted and more than a single operator $Y$ is required to globally describe the Coulomb branch.
Indeed, only four of the nine $Y_i$ operators are lifted. Continuity of the globally defined $Y$ in (\ref{eq:Y:prod:Yi}) imposes three constraints. We thus require two operators to describe the Coulomb branch. The first is the usual $Y$ that we have already introduced. The second is required to describe the novel properties in regions II and III. 
One operator that is continuous between those two regions is $\tilde{Y}_2$. However, since $Y$ itself is already continuous across these regions, a combination of $\tilde{Y}_2$ and $Y$ can also be used. We demonstrate below that the most useful definition for the second coordinate of the Coulomb branch is the combination
\begin{equation}
\tilde{Y} = \sqrt{Y \tilde{Y}_2} =\sqrt{\tilde{Y}_1 \tilde{Y}_2^2 Y_3} = \sqrt{{Y}_1' \tilde{Y}_2^2 \tilde{Y}_3}  \ .
\label{Ytildedef}
\end{equation}
This has the correct quantum numbers to be identified with a meson of the dual description\footnote{The same operator has been shown to be un-lifted on the Coulomb branch of $SO(6)$ with $F$ vectors \cite{Aharony:2013kma}.}.

The physics of these two variables can be understood as follows. In the presence of only fundamental matter, the non-perturbative contributions to the superpotential push the eigenvalues of the adjoint to be as far apart as possible. This leads to as many massless matter fields as possible with a non-vanishing adjoint and the configuration
\begin{equation}\label{Y}
Y \leftrightarrow \left( \begin{array}{cccc} \sigma \\ & 0 \\ & & 0 \\ & & & -\sigma \end{array} \right)\ .
\end{equation}
In the presence of antisymmetric tensors, however, the dynamics is somewhat different. The antisymmetric provides two doublets under the $k^\text{th}$ $SU(2)$ subgroup corresponding to the simple root $\alpha_k$. These behave like the doublets from a fundamental representation except the corresponding adjoint contributions to real masses are replaced by $(\sigma_i+\sigma_k,\sigma_i+\sigma_{k+1})$, where $i\neq k$ nor $k+1$, rather than $(\sigma_k,\sigma_{k+1})$ for fundamentals. The effective masses of the doublets provided by the antisymmetric can be ordered within a Weyl chamber similarly to the masses of the doublets from a fundamental representation. The argument leading to (\ref{Y}) can then be repeated, except the different structure of the effective real mass leads to a different unlifted direction in the Coulomb branch,
\begin{equation}
\tilde Y \leftrightarrow \left( \begin{array}{cccc} \sigma \\ & \sigma \\ & & -\sigma \\ & & & -\sigma \end{array} \right) \ .
\end{equation}
When both anti-symmetric tensors and fundamentals are present, the un-lifted Coulomb branch is two dimensional and can be parametrized generically as
\beq
\label{eq:breaks}
\left(
\begin{array}{cccc}
\sigma_1&&&\\
&\sigma_2&&\\
&&-\sigma_2&\\
&&&-\sigma_1
\end{array}
\right)
\quad{\rm with}\quad
\left\{
\begin{array}{lll}
Y &\leftrightarrow & {\rm diag} \big(\sigma_1-\sigma_2,0,0,\sigma_2-\sigma_1\big)\\
\\
\tilde{Y}& \leftrightarrow & {\rm diag} \big(\sigma_2,\sigma_2,-\sigma_2,-\sigma_2\big)\\
\end{array}
\right.
.
\eeq
Since regions I and IV have a zero mode structure similar to the case with only fundamental matter, those regions are always described by the operator $Y$. In regions II and III, however, the antisymmetric wants to align the adjoint into the $\tilde{Y}$ direction. 

The arguments presented here are semi-classical. Depending on the specific matter content of the theory, there may be important additional dynamical effects that modify the semi-classical picture~\cite{Poppitz:2013}. 
In the next section we investigate the dynamics of the $SU(4)$ theory with one and two antisymmetric tensors. We show that in theories with one anti-symmetric, only the direction $Y$ is present, suggesting that $\tilde Y$=0. Both $Y$ and $\tilde{Y}$ are needed to describe the dynamics of theories with two antisymmetric tensors. We show this both by considering the dimensional reduction of 4D theories with one and two antisymmetric tensors, and by studying the decoupling of an antisymmetric from a 3D theory with two antisymmetrics, which sets $\tilde{Y}=0$.

As already mentioned above, both $Y$ and $\tilde Y$ are the low energy limits of monopole operators. Such an identification has multiple subtleties. We address them next.


\subsection{Dirac quantization condition of the monopole operators}

Monopoles are subject to the Dirac quantization condition which restricts the minimal magnetic charge that can appear in a theory. This, in turn, restricts the form of the monopole operators that describe the un-lifted Coulomb branch directions.
If there are unbroken non-Abelian subgroups, then magnetic charges smaller than the na\"ive minimal value might be allowed, as long as the monopole carries non-Abelian magnetic charge that is screened at $r\sim \Lambda_\text{confine}^{-1}$. This is analogous to the well known case of GUT monopoles, see e.g.~\cite{Preskill:1984gd}.
We therefore need to check that both $Y$ and $\tilde{Y}$ are the appropriate ``minimal'' monopole operators allowed by Dirac quantization. In particular, the definition of $\tilde{Y}$ might raise concerns since the square root would naively imply the presence of a monopole with a fractional charge. 

As it is indicated in 
(\ref{eq:breaks}), $Y$ and $\tilde{Y}$ are associated to the breaking $SU(4)\to SU(2)\times U(1)_1\times U(1)_2$ and $SU(4)\to SU(2)\times SU(2) \times U(1)_3$ respectively. We can explicitly identify the generators of each one of the unbroken $U(1)$s
\begin{align}
Y:&\qquad Q_1=
\begin{pmatrix}
1 & & &\\
& 0 & &\\
& & 0 &\\
& & & -1
\end{pmatrix}
&
Q_2&= 
\begin{pmatrix}
1 & & &\\
& -1 & & \\
& & -1 & \\
& & & 1
\end{pmatrix},
\\\label{Q3}
\tilde{Y}:&\qquad Q_3=
\begin{pmatrix}
1 & & & \\
& 1 & & \\
& & -1 & \\
& & & -1
\end{pmatrix},
\end{align}
where $Q_i$ generates the corresponding $U(1)_i$. While the $Y$ operator is associated to a monopole operator with a minimal Dirac charge\footnote{
The magnitude of the magnetic charge depends on the normalization of the U(1) generators. However, this normalization also affects the minimal allowed electric charge so that a more meaningful quantity is the magnetic charge in units of the minimal electric charge. This quantity is independent of the normalization of the generators.
} 
under $U(1)_1$, the square root in the definition of $\tilde{Y}$ in (\ref{Ytildedef}) suggests that this operator corresponds to half of the minimal charge and therefore is not allowed~\cite{Aharony:2013kma}. 

However, as described above, in the presence of unbroken non-Abelian subgroups, the Dirac quantization condition can be more subtle. We argue that Dirac quantization is, in fact, obeyed by the state corresponding to the operator $\tilde{Y}$. The key aspect is that while the symmetry breaking pattern corresponding to $\tilde{Y}$ is locally $SU(4)\to SU(2)\times SU(2)\times U(1)$, there is a non-trivial identification between a discrete $U(1)$ rotation and one of the elements of the $\mathbb Z_2\times \mathbb Z_2$ center of $SU(2)\times SU(2)$. From the explicit definition above, it is clear that $\exp{(i\pi Q_3)}$ coincides with the  ${\rm diag}(-1,-1-1,-1)$ element of the center of $SU(2)\times SU(2)$. In such a case, one need not go around the entire $U(1)$ factor to obtain a closed loop; the loop can be closed by going halfway around and then closing through the $\mathbb Z_2$ center. In this case the monopole picks up a discrete magnetic $\mathbb Z_2$ charge under the $SU(2)\times SU(2)$ non-Abelian group. For cases like this, the proper formulation of the Dirac quantization condition has been given by Preskill~\cite{Preskill:1984gd}: a magnetic charge is allowed if one can combine the charge matrix with some of the diagonal non-Abelian generators to obtain the full magnetic charge matrix $M$ which has only integer elements. In our case the proper choice is 
\begin{equation}\label{genmono}
M= \frac{1}{2}Q_3 +T_3^{(1)}+ T_3^{(2)} = \left( \begin{array}{cccc} 1 \\ & 0 \\ & & 0 \\ & & & -1\end{array} \right)\ .
\end{equation}
This is in accordance with our definition that the operator $\tilde{Y}$ carries charge $1/2$ under $Q_3$.

A more physical way of describing the above argument is to only consider the theory at large distances. When considering the breaking pattern described by $\tilde{Y}$, an $SU(4)$ fundamental decomposes into a $(2,1)\oplus(1,2)$ under the unbroken $SU(2)\times SU(2)$. At low energies, both of the $SU(2)$s are still strongly interacting and the only physical states are composite objects. This requires at least two doublets originating from two fundamentals under $SU(4)$.  The minimal charge under the $U(1)_3$ is thus double the one na\"ively given by a fundamental, and therefore the minimal Dirac charge at low energy is in fact a half of the na\"ive result. This provides an additional explanation for the presence of the square root in the definition of $\tilde{Y}$.
This is similar to the argument that is usually given for why the fractional electric charges of quarks do not forbid the appearance of a minimally charged Dirac monopole in GUT extensions of the Standard Model. It is also clear that such an argument cannot be applied for the breaking associated with $Y$: when considering a fundamental, during the breaking of $SU(4)$ to $SU(2)\times U(1)\times U(1)$, one component of the fundamental is only charged under the U(1) with the minimal charge. The possible presence of this state leads to the usual Dirac quantization without any subtleties.

\section{Duals of 3D theories with antisymmetrics}
\label{check:duals}

We now discuss the dynamics of the 3D $SU(4)$ theories with antisymmetric tensors. First we consider a 3D $\mathcal{N}=2$ SUSY gauge theory with two antisymmetric tensors and two flavors. The corresponding theory in 4D has three flavors and the anomaly-free symmetries:
\begin{align}
\begin{tabular}{l|c|cccccc}
&$SU(4)$&$SU(2)$&$SU(3)_L$&$SU(3)_R$&$U(1)$&$U(1)^\prime$&$U(1)_R$\\
\hline
$A$&$\antin$&$\square$&$\I$&$\I$&0&-3&$0$  \vphantom{$\sqrt{\anti}$}\\
$Q$&$\square$&$\I$&$\square$&$\I$&1&2&$\frac{1}{3}$\\
$\overline{Q}$&$\overline{\square}$&$\I$&$\I$&$\overline{\square}$&-1&2&$\frac{1}{3}$
\end{tabular}
\end{align}
It is well-known that this theory is s-confining in 4D \cite{Csaki:1996zb,Csaki:1996sm,Csaki:1998th}: it is dual to a theory of gauge singlets with the following non-trivial superpotential, which is smooth everywhere including the origin:
\begin{align}\label{WSU(4)}
W_{\rm dyn}=\frac{1}{\Lambda^7}\big(T^2M_0^3-12 T H \bar{H}M_0-24 M_0M_2^2-24 H \bar{H} M_2\big)
\end{align}
where the composite fields have the following charge assignment under the global symmetries:
\begin{align}
\begin{tabular}{c|c|cccccc}
&$SU(4)$&$SU(2)$&$SU(3)_L$&$SU(3)_R$&$U(1)$&$U(1)^\prime$&$U(1)_R$\\
\hline
$M_0=Q\bar{Q}$&$\I$&$\I$&$\square$&$\overline{\square}$&0&4&$\frac{2}{3}$ \vphantom{$\sqrt{\af}$}\\
$M_2=QA^2\bar{Q}$&$\I$&$\I$&$\square$&$\overline{\square}$&0&-2&$\frac{2}{3}$\\
$H=AQ^2$&$\I$&$\square$&$\overline{\square}$&$\I$&2&1&$\frac{2}{3}$\\
$\bar{H}=A\bar{Q}^2$&$\I$&$\square$&$\I$&$\square$&-2&1&$\frac{2}{3}$\\
$T=A^2$&$\I$&${\tiny\Yvcentermath1 \yng(2)}$&$\I$&$\I$&0&-6&$0$
\end{tabular}
\end{align}

\subsection[3D duality for $SU(4)$ with 2 $A$ and 2 ($Q+\bar Q$)]{3D duality for $SU(4)$ with 2 $\antin$ and 2 ($\f+\af$)}\label{sec:anti}

Based on the rules established in~\cite{Aharony:2013dha} one can obtain a dual pair on $\mathbb R^3\times S^1$ by adding to the superpotential an $\eta Y$ term generated by KK monopoles on both sides of the duality\footnote{In an s-confining theory, such a term is only added to the electric descriptions since there are no monopoles in the dual (low energy) theory.}.  In the presence of antisymmetric tensors, this statement requires more care.
In a KK monopole background, matter in a generic representation can provide extra zero modes and prevent a contribution to the superpotential. It is known that fundamentals have no zero modes in the KK monopole background but an antisymmetric generically does. Yet this only happens for $SU(N)$ with $N>4$ but not for $SU(4)$. This statement can be checked in two equivalent ways.

\begin{enumerate}
\item One can explicitly compute the number of the antisymmetric zero modes in the KK monopole configuration using the appropriate index theorem on $\mathbb R^{3}\times S^1$ \cite{Nye:2000it} which has been very nicely fleshed out recently by Poppitz and Unsal \cite{Poppitz:2008it}. The result of the index theorem counting is that the antisymmetric has no KK monopole zero modes in any of the four regions. 
\item One can exploit the important property of the ordinary $SU(N)$ instanton: compactified on a circle it can be thought of as a composite of the $(N-1)$ fundamental instanton/monopoles and the KK monopole \cite{Lee:1997mo,Lee:1998mo,Lee:1998ca,Kraan:1998pi,Kraan:1998cal,Kraan:1998mon}. Thus the total number of zero modes in $\mathbb R^3\times S^1$ for all $N$ independent monopole solutions (instanton/monopoles plus the KK monopole) should match the number of zero modes of the one 4D instanton given by the Atiyah-Singer index theorem. Therefore the number of zero modes in the KK monopole for a given representation is given by the difference of the 4D instanton zero modes and the sum of zero modes in the $(N-1)$ independent 3D instanton/monopoles. The latter can be obtained directly from the 3D Callias index theorem. We can see from (\ref{CoulombSU(4)}) that in each of the four regions of the Coulomb branch, the total number of zero modes of the antisymmetric matches the number of zero modes in the four dimensional instanton solution. Thus, for $SU(4)$ the antisymmetric does not have any zero modes in the KK monopole which, in turn, generates a non-trivial superpotential in each of the 4 regions of the Coulomb branch. 
\end{enumerate}

To obtain the pure 3D dual pair one can remove the $\eta Y$ term by introducing a real mass term for one flavor.  This can be most easily carried out on the dual pair by weakly gauging the $\big[SU(3)_L\times SU(3)_R\big]_D\times U(1)$ subgroup of the global symmetries. One can then introduce constant scalar backgrounds in the $U(1)$ and along the $\lambda_8$ direction of the diagonal $SU(3)$ in such a way that the first two generations remain massless while the third generation quarks $Q_3$ and $\bar{Q}_3$ acquire real masses $m_\mathbb{R}$ and $-m_\mathbb{R}$ respectively.  Such a background configuration breaks the global $SU(3)_L\times SU(3)_R\times U(1)$ to $SU(2)_L\times SU(2)_R\times U(1)_1\times U(1)_2$. Furthermore in the limit $m_R\to \infty$, the $\eta Y$ term decouples \cite{Aharony:2013dha}. Thus on the electric side we flow to an $SU(4)$ gauge theory with no superpotential and the following matter content:
\begin{align}\label{contentSU(4)}
\begin{tabular}{l|c|ccccccc}
&$SU(4)$&$SU(2)$&$SU(2)_L$&$SU(2)_R$&$U(1)_1$&$U(1)_2$&$U(1)^\prime$&$U(1)_R$\\
\hline
$A$&$\antin$&$\square$&$\I$&$\I$&0&0&-3&$0$\vphantom{$\sqrt{\anti}$}\\
$Q$&$\square$&$\I$&$\square$&1&1&0&2&$\frac{1}{3}$\\
$\overline{Q}$&$\overline{\square}$&$\I$&$\I$&$\overline{\square}$&0&-1&2&$\frac{1}{3}$
\end{tabular}
\end{align}
It is just a 3D $SU(4)$ theory with two antisymmetric tensors and two flavors. 

In order to find the effect of these background scalars on the dual we need to carefully identify their effects on the composites. We find that the following dual fields remain massless:
\begin{eqnarray}
M_0^{\phantom{0}ia}
&\quad\to\quad&
M_0^{\phantom{0}ia}\ i,a=1,2,\quad M_0^{\phantom{0}33}\\
M_2^{\phantom{2}ia}
&\quad\to\quad&
M_2^{\phantom{2}ia}\ i,a=1,2,\quad M_2^{\phantom{0}33}\\
H^{Ii}
&\quad\to\quad&H^{I3}\\
\bar{H}^{Ia}
&\quad\to\quad&\bar{H}^{I3}
\end{eqnarray}
As expected, almost all composites containing a third generation quark or antiquark decouple, except for $M_0^{33}$ and $M_2^{33}$: since the real masses of the quark and the antiquark are of opposite sign, these fields remain in the spectrum. Using the field identification $H^{I3}\equiv h^I $,  $\bar{H}^{I3}\equiv \bar{h}^I$ , $M_0^{\phantom{0}33}\equiv \tilde{M}_0$,  $M_2^{\phantom{0}33}\equiv \tilde{M}_2$ we find the low-energy matter content of the dual theory to be
\begin{align}\label{tableSU(4)}
\begin{tabular}{c|ccccccc}
&$SU(2)$&$SU(2)_L$&$SU(2)_R$&$U(1)_1$&$U(1)_2$&$U(1)^\prime$&$U(1)_R$\\
\hline
$M_0$&$\I$&$\square$&$\overline{\square}$&1&-1&4&$\frac{2}{3}$\vphantom{$\tilde{M}_0$}\\
$M_2$&$\I$&$\square$&$\overline{\square}$&1&-1&-2&$\frac{2}{3}$\\
$h$&$\square$&$\I$&$\I$&2&0&1&$\frac{2}{3}$\\
$\overline{h}$&$\square$&$\I$&$\I$&0&-2&1&$\frac{2}{3}$\\
$T$&${\tiny\Yvcentermath1 \yng(2)}$&$\I$&$\I$&0&0&-6&0\\
\hline
$\tilde{M}_0$&1&1&1&-2&2&4&$\frac{2}{3}$\vphantom{$\tilde{M}_0$}\\
$\tilde{M}_2$&$\I$&$\I$&$\I$&-2&2&-2&$\frac{2}{3}$
\end{tabular}
\end{align}
with the superpotential (we don't explicitly write an overall scale needed on dimensional grounds)
\begin{align}\label{lowSU(4)}
W_{\text{dyn}}= \tilde{M}_0\Big(3 T^2 \det M_0-12 Th\overline{h}-24\det M_2\Big)+\tilde{M}_2\Big(2 M_0 M_2+h \overline{h}\Big)
\end{align}

Most of the chiral operators in (\ref{tableSU(4)}) and (\ref{lowSU(4)}) are easily identified as meson operators of the electric theory (\ref{contentSU(4)}), yet for both $\tilde{M}_0$ and $\tilde{M}_2$ such identification fails. This is not a new feature. In the $SU(N)$ SQCD case, one of the meson operators of the magnetic theory is identified with the chiral operator describing the unlifted region of the Coulomb branch of the electric theory. Such identification is explained in more detail in \cite{Aharony:2013dha}. We claim that similar dynamics takes place in the present case and that the mesons $\tilde{M}_0$ and $\tilde{M}_2$ are identified with the Coulomb branch moduli $Y$ and $\tilde Y$ of the electric theory:
\beq
\tilde{M}_0 \to Y,\qquad \tilde{M}_2\to \tilde{Y}\ .
\eeq

\subsection{Consistency checks of the duality}

\subsubsection{Matching charges of the Coulomb branch operators}

The first check on the proposed duality is simply a matching of the quantum numbers of the Coulomb branch operators. Using the zero mode counting summarized in table (\ref{CoulombSU(4)}) together with the quantum numbers of the elementary fields in 
  (\ref{contentSU(4)}) we can explicitly compute the charge assignment for the Coulomb branch operators, which correspond to the global charges carried by the fermionic zero modes in a given one-instanton background. The resulting charges are

\begin{align}
\begin{tabular}{c|cccc}
&$U(1)_1$&$U(1)_2$&$U(1)^\prime$&$U(1)_R$\\
\hline
\noalign{\vskip 1mm}   
$Y_1$&-2&2&4&$\frac{14}{3}$\\
$Y_2$&0&0&0&-2\\
$Y_3$&0&0&0&-2\\
$\tilde{Y}_1$&0&0&12&2\\
$\tilde{Y}_2$&-2&2&-8&$\frac{2}{3}$\\
$\tilde{Y}_3$&0&0&12&2\\
$Y'_1$&0&0&0&-2\\
$Y'_2$&0&0&0&-2\\
$Y'_3$&-2&2&4&$\frac{14}{3}$ \\ \hline \hline \noalign{\vskip 1mm}  
$Y=\prod_i Y_i$ & -2 & 2 & 4 & $\frac{2}{3}$ \\ 
$\tilde{Y}= \sqrt{Y \tilde{Y}_2}$ & -2 & 2 & -2 & $\frac{2}{3}$ \\
\end{tabular}
\end{align}
We can see that the charges of $Y,\tilde{Y}$ indeed match with those of $\tilde{M}_0, \tilde{M}_2$ from table (\ref{tableSU(4)}).

\subsubsection{Matching the quantum constraints on a circle}

The second consistency check is to reproduce the dynamics of the theory on a circle ($\mathbb R^3\times S^1$). To obtain the description on the circle from 3D theories, we need to add the contribution of the KK monopole\footnote{The KK monopole contribution can be lifted by matter zero modes. But, as described in Section~\ref{sec:anti}, in the presence of antisymmetric tensors, this only happens for $SU(N)$, with $N>4$.} (see~\cite{Lee:1997mo,Lee:1998mo,Lee:1998ca} or Appendix~\ref{app:dim:reduction}). 
In our model, a calculation \`a la Polyakov gives the KK monopole contribution of the form
\beq
W_\text{KK} = \eta Y \sim \Lambda^8 Y \ .
\eeq
This modifies the superpotential to
\begin{align}\label{IR3d}
W_{\text{dyn}}=Y\Big(3 T^2 \det M_0-12 Th\overline{h}-24\det M_2\Big)+\eta Y + \tilde{Y}\Big(2 M_0 M_2+h \overline{h}\Big)
\end{align}
On the other hand, the theory on the circle can be obtained directly from the 4D description by compactifying one of the spatial directions both for the electric and magnetic sides of the 4D duality. It is well-known that the physics of the 4D $SU(4)$ gauge theory with 2 flavors and 2 antisymmetrics is described by a set of gauge invariant fields satisfying two constraints, one of which is quantum modified while the other is not~\cite{Csaki:1996zb,Csaki:1996sm,Csaki:1998th}. These constraints can be captured by the superpotential 
\begin{align}
\label{IR4d}
 W=\lambda\left(3T^2 \det M_0-12Th\bar{h}-\right.&\left.24 \det M_2-\Lambda^8\right)+\mu\left(2M_0M_2+h \bar{h}\right)
\end{align}
where $\lambda$ and $\mu$ are Lagrange multipliers enforcing the constraints.
A comparison of (\ref{IR3d}) and (\ref{IR4d}) suggests an identification of the Lagrange multipliers with the Coulomb branch moduli, $\lambda\to Y$ and $\mu\to \tilde{Y}$. Indeed, integrating out $Y$ and $\tilde Y$ from (\ref{IR3d}) reproduces the constraints imposed by (\ref{IR4d}).

\subsection[Duality for $SU(4)$ with $A$ and $3(Q+\bar Q)$]{Duality for $SU(4)$ with $\antin$ and $3(\f+\af)$}\label{sec:oneanti}

We can also connect the theory investigated above to the sequence obtained from $SU(4)$ with a single antisymmetric tensor and flavors. The 4D s-confining version of this sequence with one antisymmetric is the model with 4 flavors. Carrying through the steps of compactification on a circle and adding a real mass we arrive the 3D s-confining version of the theory with a single antisymmetric and three flavors, given by the superpotential
\beq\label{oneanti}
  W=Y\left(T\det\,M + H M \bar{H}\right)
\eeq
where $T=A^2, M = Q\bar{Q}, H= AQ^2$ and $\bar{H}=A\bar{Q}^2$.  Note, that this theory has a single Coulomb branch operator $Y$ appearing in its low energy dynamics. Giving a complex mass $\mu$ to one more flavor from this theory we obtain a low energy description with a quantum modified constraint given by the superpotential
\begin{equation}
\label{oneantiQM}
W=\lambda \left[ Y(T \det M + H\bar{H}) - \mu\right]
\end{equation}
As a consistency check we can verify whether this quantum modified constraint is obtained when a holomorphic mass term $\mu A_2^2$ is added to one of the antisymmetric fields of the model discussed in section  \ref{sec:anti}:
\begin{align}\label{integrateout}
W_{\text{dyn}}=Y\Big(3 T^2 \det M_0-12 Th\overline{h}-24\det M_2\Big)+ \tilde{Y}\Big(2 M_0 M_2+h \overline{h}\Big) + \mu T_{22}\ .
\end{align}
Indeed the $T_{22}$ equation of motion will provide precisely the constraint in (\ref{oneantiQM}) while together with the equations of motion for the other fields containing $A_2$ the entire superpotential will be set to zero. In particular, any reference to the variable $\tilde{Y}$ disappears (without having to take the equation of motion for $\tilde{Y}$). This suggests that the Coulomb branch dynamics in regions II and III changes when one goes from two antisymmetrics to a single one: in the latter case the fundamental will dominate the dynamics and a single variable will be sufficient to describe the entire Coulomb branch. We propose a tentative explanation for it in the next section.

\subsection{Low energy description on the Coulomb branch}
Another important check of the duality involves the analysis of the low energy physics on the Coulomb and Higgs branches of the theory. While we leave the detailed analysis for future work \cite{Csaki:2014new}, let us point out some qualitative features of this regime. On the electric side, at large $\tilde Y$ the $SU(4)$ gauge group is broken to an $SU(2)\times SU(2)\times U(1)$. Let us concentrate on the dynamics of the $SU(2)\times SU(2)\cong SO(4)$ group. Components of the fundamental fields charged under this group obtain large real masses and are expected to decouple from the low energy physics while the light degrees of freedom in the anti-symmetric tensors transform as vectors of the $SO(4)$. 

On the magnetic side there is no gauge group and thus no Coulomb branch. $\tilde Y$ only appears as a chiral superfield, therefore we expect that the effect of decoupling the fundamentals will be taken care of by the non-trivial superpotential in (\ref{lowSU(4)}). This expectation is in fact accurate. At large $\tilde Y$, all composite fields containing electric fundamental degrees of freedom acquire a mass through (\ref{lowSU(4)}) and become heavy. Integrating them out we find that the IR degrees of freedom are $Y$, $\tilde Y$, and $T$, while the superpotential vanishes. This also agrees with the results of \cite{Aharony:2011ci} where it was found that in the $SO(4)$ theory with two vectors, the IR superpotential vanishes, the classical moduli space is not modified quantum mechanically, but the classical singularity at the origin is smoothed out. The massless degrees of freedom in these two theories match once the dependence of the $SO(4)$ coupling on $\tilde Y$ is taken into account.

It is also interesting to consider the Coulomb branch dynamics of a theory with a single antisymmetric. 
In the electric theory, along the $\tilde Y$ direction, an effective superpotential is generated for the modulus $Y_{SO}$ of the unbroken $SO(4)$ subgroup of the form~\cite{Aharony:2011ci}
\begin{equation}
 W=\frac{1}{Y_{SO}^2 T}\,.
\end{equation}
Matching $Y_{SO}$ to the operators of the full $SU(4)$ theory one can see that the fields are pushed towards the $Y$ branch while the $\tilde Y$ direction is lifted. This resolves a puzzle we encountered in the previous subsection -- while the counting of the instanton-monopole zero modes suggest a possibility of two unlifted moduli, the correct magnetic description (\ref{oneanti}) found through the reduction of the 4D theory involves only $Y$. Such considerations are independent of the number of flavors present in the theory. It is therefore reasonable to assume that the presence of  a single unlifted Coulomb branch direction parametrized by $Y$ is a generic feature of $SU(4)$ theories with a single antisymmetric tensor.

\section{Classification of 3D s-confining theories}

We can now briefly outline how the previous discussion generalizes to all three dimensional s-confining theories.
We only make preliminary comments and leave a general classification of 3D s-confinement to subsequent work~\cite{Csaki:2014new}.

Anomaly matching techniques cannot be used in 3D\footnote{%
In odd dimensions, gauge invariance can require the addition of a classical Chern-Simons term which breaks parity. This is referred to as the parity anomaly \cite{Niemi:1983rq, Redlich:1983dv,Redlich:1983kn}.
}, yet it has been argued in \cite{Witten:1999in,Intriligator:2013lca} that a necessary condition for confinement in 3D is that the Witten index  ${\rm Tr}(-{\bf 1})^F$ equals 1.
The Witten index \cite{Witten:1982sb} for an $SU(N)$ gauge theory with Chern-Simons number $k$ and generic matter content in the ${\bf r}_f$ representation has been recently computed to be \cite{Intriligator:2013lca,Smilga:2013in}:
\begin{align}\label{witind}
{\rm Tr}(-{\bf 1})^F&=\frac{(k'+N-1)!}{(N-1)!k'!}, 
& 
k'&=|k|-T_2(\text{adj.})+\sum_f T_2({\bf r}_f)
\end{align}
where the sum is over the matter fields in representation ${\bf r}_f$, and $T_2({\bf r}_f)$ is the Dynkin index normalized to $1/2$ for fundamentals. (\ref{witind}) implies that 3D s-confining theories satisfy~\cite{Witten:1999in,Intriligator:2013lca}:
\begin{align}\label{cond}
k'=0\quad\Rightarrow\quad|k|=T_2(\text{adj.})-\sum_fT_2({\bf r}_f).
\end{align}
Let us now focus on the case with vanishing Chern-Simons index, $k=0$.
In such a case, (\ref{cond}) reduces to the condition for a 4D theory to exhibit a quantum modified constraint. Such theories can often be obtained by decoupling one flavor from 4D s-confining theories.
In order to derive an explicit expression for the superpotential 3D s-confining theories, one must first dimensionally reduce the 4D s-confining theory which has an extra flavor compared to its 3D s-confining partner. 
Next one must make a real mass deformation to remove one flavor and set the Witten index to 1. This also removes the $\eta Y$ term from the electric superpotential. During this process, the IR description of the 3D theory is always under control and one is be able to explicitly write down the superpotential.

Based on the examples presented in this paper as well as the original case of 3D SQCD~\cite{Aharony:1997bx,Aharony:2013dha,deBoer:1997kr}, we expect the following structure to emerge. The matter content of the s-confining 3D theories  correspond to 4D models with quantum moduli spaces described by several constraints among the gauge invariant fields, one of which is quantum modified, while the others are not. We expect that the number of unlifted Coulomb branch directions should match the number of constraints present in the 4D theory. In other words, the surviving Coulomb branch operators are identified with the Lagrange multipliers of the 4D theory once these are promoted to dynamical fields. 

The $SU(4)$ model with $2\ \antim +2(\f+\af)$ in this paper is part of a  sequence of theories $SU(N)\ \left(\antim+\aanti\right)+2(\f+\af)$ which satisfiy (\ref{cond}) and are expected to s-confine. The 4D description is well known, and the number of constraints are expected to indicate the number of unlifted Coulomb branch operators:
\begin{eqnarray}\label{QMC1}
SU(2N) & {\rm with}&\left(\antim+\aanti\right)+2\left(\f+\af\right)\quad\to\quad N\ {\rm constraints}\\\label{QMC2}
SU(2N+1) & {\rm with}& \left(\antim+\aanti\right)+2\left(\f+\af\right)\quad\to\quad N\ {\rm constraints}
\end{eqnarray}
For $SU(4)$, $SU(5)$ and $SU(6)$ the Coulomb branch operators corresponding to the directions of the various constraints can be identified in the following way:
\begin{align}
\begin{array}{cllllllll}
SU(4) & 
Y_4&\to& {\rm diag}(\sigma ,0,0,-\sigma ) 
& \tilde{Y}&\to& \sqrt{YY_2} &\to& {\rm diag}(\sigma ,\sigma , -\sigma , -\sigma ) 
\\
SU(5) 
& Y_5&\to& {\rm diag}(\sigma ,0,0,0,-\sigma ) 
& \tilde{Y}_5&\to& \sqrt{YY_2Y_3} &\to& {\rm diag}(\sigma ,\sigma ,0, -\sigma , -\sigma ) 
\\
SU(6) 
& Y_6&\to& {\rm diag}(\sigma ,0,0,0,0,-\sigma ) 
& \tilde{Y}_6&\to& \sqrt{YY_2Y_3Y_4} &\to& {\rm diag}(\sigma ,\sigma ,0,0, -\sigma , -\sigma ) 
\\ 
&
&
& 
&  
\hat{Y}_6&\to& (\tilde{Y}_1^2 Y_3)^\frac{1}{3} &\to& {\rm diag}(\sigma ,\sigma ,\sigma,-\sigma, -\sigma , -\sigma )
\end{array}
\end{align}
with an obvious generalized pattern for even and odd $N$.  The quantum numbers of these operators exactly match the quantum numbers of the ``extra'' meson operators which remain massless after the real mass deformation and suggest that  these Coulomb branch operators are identified with the extra mesons.

We now argue for the validity of these operators.
All of these operators correspond to monopole configurations and satisfy the Dirac quantization condition. The $SU(5)$ case is very similar to $SU(4)$ with respect to these monopole operators. In fact, it can be checked that the direction described by $\tilde{Y}_5$ breaks $SU(5)\to \big(SU(2)^2\times U(1)^2\big)/\mathbb{Z}_2$. Half-integer charged monopoles are allowed and therefore justify the square root in the definition of the monopole operator. A similar discussion applies for $\tilde{Y}_6$. 

New features arise when considering $\hat Y_6$ operator in an $SU(6)$ theory. Along the this direction, the symmetry breaking is locally $SU(6)\to SU(3)^2\times U(1)$, with the $U(1)$ direction generated by
\begin{align}
Q_6 = \text{diag}\left(\frac 13, \frac 13, \frac 13, -\frac 13, -\frac 13, -\frac 13\right).
\end{align}
There are multiple identifications among the $U(1)$ orbit and the $\mathbb Z_3\times\mathbb Z_3$ center of $SU(3)\times SU(3)$, thus modifying the global structure of the breaking into $SU(6)\to \big(SU(3)^2\times U(1)\big)/\mathbb Z_3$. We therefore expect fractionally charged monopole, in particular with $1/3$ of the minimal Dirac charge. 

\section{Conclusions}

In this paper we initiated the study of the dynamics of 3D $\mathcal N=2$ supersymmetric theories with matter fields in generic representations.
Already in the simplest extension of 3D SQCD obtained by adding antisymmetric matter, we have found interesting new features.

While previously studied theories have at the most one unlifted Coulomb branch direction, we have shown that the physics is very different when antisymmetric tensors are added to the theory because multiple directions along the Coulomb branch could remain unlifted. We have examined the case of $SU(4)$ with flavors and one or two anti-symmetric tensors.
By performing a careful analysis of fermionic zero modes in monopole backgrounds 
and matching to the dynamics of 4D theories,
we showed that theories with two antisymmetric tensors have two unlifted Coulomb branch directions. We also identified the additional Coulomb branch modulus $\tilde{Y}$ with a fractionally charged monopole operator.

We analyzed the flow to a 3D s-confining theory with two flavors and two antisymmetric tensors by performing a dimensional reduction of 4D s-confining dualities and decoupling a flavor through a real mass deformation. The quantum numbers of the two Coulomb branch operators, $Y$ and $\tilde Y$, exactly match the quantum numbers of certain meson operators which naturally appear in the superpotential obtained by the dimensional reduction. This is strong evidence that our Coulomb branch analysis is correct. Furthermore we identify both $Y$ and $\tilde Y$ as monopole operators associated to particular monopole configurations in the UV. In particular $\tilde Y$ is associated with a fractionally charged monopole and is allowed because of the non-trivial global topology of the unbroken group along the particular Coulomb branch direction described by $\tilde Y$.

We provide multiple checks for our dualities. The 3D magnetic dual correctly reproduces the behavior expected on $\mathbb R^3 \times S^1$, which can be obtained by compactifying the 4D $SU(4)$ theory with two antisymmetric tensors and two flavors. 
It is known that in 4D, this theory is described in the IR by a quantum modified and a classical constraint.
From a 3D perspective, the quantum modification of one of the classical constraints arises from the $\eta Y$ operator generated by the dynamics of the KK monopole.
%
A similar analysis in $SU(4)$ s-confining theories with a single antisymmetric suggests that only a single Coulomb branch direction remains in such a theory, while the $\tilde Y$ direction is lifted. While a detailed explanation of this phenomenon is beyond the scope of this paper, we provided a tentative explanation of such dynamics looking at the ADS-like superpotential generated along the $\tilde Y$ direction. 

Finally we also presented initial comments on the description of more general 3D s-confining theories. The details of the general analysis is left for future work.

\section*{Acknowledgements}

The authors thank
Philip Argyres,
Ken Intriligator,
Erich Poppitz,
Arvind Rajaraman,
Mithat Unsal,
and
Brian Willett
for useful comments and discussions. 
We are especially grateful to Ofer Aharony for discussions and many important comments and suggestions on earlier drafts of this paper. C.C. and M.M. thank the Mainz Institute for Theoretical Physics (MITP) for its hospitality while this paper was concluded.
C.C.\ and M.M.\ are supported in part by the NSF grant PHY-1316222. Y.S.\ and P.T.\ are supported in part by the NSF grant grant PHY-1316792. J.T.\ is supported in part by the Department of Energy under grant DE-FG02-91ER406746.

\appendix

\section*{Appendix}

\section{Elements of $\mathcal N=2$ SUSY in Three Dimensions}
\label{app:notation}

3D $\mathcal N=2$ supersymmetry can be obtained by dimensional reduction of familiar 4D $\mathcal{N}=1$ SUSY.
We review aspects of $\mathcal N=2$ SUSY in 3D and introduce the notation used in this manuscript. 
%
%
Most of the material presented here follows the treatment in \cite{Affleck:1982as,Aharony:1997bx,deBoer:1997kr,Intriligator:2013lca}


\subsection{3D Spinors and the $\mathcal N=2$ SUSY Algebra}

A convenient representation of the Clifford algebra in 3D with metric $\eta_{ij} = (-,+,+)$ is
\begin{align}
\gamma^{i=0,1,2}_{\alpha,\beta} = (i\sigma_2,\sigma_3,\sigma_1).
\label{app:SUSY:gamma}
\end{align}
Note that the generators of the Lorentz group, $S_{ij} = \epsilon_{ijk}\frac 12 \gamma^k$, are pure imaginary and thus generate a real group. The fundamental fermion representation in (2+1)-dimensions is therefore a 2-component Majorana fermion, $\psi_M$. Under parity, this transforms as 
\begin{align}
P:\;\; \psi_M \to \pm \gamma_1 \psi_M. 
\label{eq:SUSY:3D:parity}
\end{align}
As usual, spinor indices are contracted, raised and lowered with $\epsilon^{\alpha\beta}$ or $\epsilon_{\alpha\beta}$.


From the usual 4D $\mathcal N=1$ algebra, the 3D $\mathcal N=2$ SUSY algebra is
\beq\label{SUSY:algebra}
\left\{Q_\alpha,\bar{Q}_\beta\right\}=2 \gamma_{\alpha\beta}^i P_i+2i \epsilon_{\alpha\beta}Z,
\eeq
where the central charge $Z$ is identified with momentum along $x^{\mu=2}$ in the 4D picture.  The realization of (\ref{SUSY:algebra}) as differential operators acting on superspace follow from the 4D formalism. From this one may read off the 3D $\mathcal N=2$ SUSY multiplets. 

\subsection{Chiral superfields}

In 4D, a chiral superfield $Q$ may be written as
\beq
Q=\phi_Q+\theta \psi+\theta^2 F
\eeq
where $\phi_Q$ is a complex scalar, $\psi$ is a Weyl fermion which decomposes into two independent real Majorana fermions in 3D, and $F$ is an auxiliary field.

\subsection{Vector superfields and the Coulomb branch}

Similarly, a 4D vector superfield may be written as
\beq\label{app:SUSY:vec}
V=-i\theta\bar{\theta}\sigma-\theta\gamma^i\bar{\theta}A_i+i\theta^2\bar{\theta}\lambda-i\bar{\theta}^2\theta\lambda+\frac{1}{2}\theta^2\bar{\theta}^2D,
\eeq
where we have explicitly separated the 3D vector $A_i$ from the gauge scalar $\sigma \sim A_3$ and the $\gamma^i$ are defined in (\ref{app:SUSY:gamma}). 
Unlike the chiral superfield, the 3D $\mathcal{N}=2$ vector superfield differs from its 4D $\mathcal N=1$ counterpart in that it carries components which may acquire vacuum expectation values (VEVs) that form the Coulomb branch of the moduli space.

The scalar component $\sigma$ may acquire a VEV which breaks the gauge group to its maximal abelian subgroup. For $SU(N)$, $\langle\sigma\rangle\neq0$ induces the breaking $SU(N)\to U(1)^{N-1}$. 
This, however, is not the only scalar that may acquire a background value. A 3D vector carries a single propagating degree of freedom and may be dualized into a scalar, $\gamma$. 
For simplicity, consider a $U(1)$ theory---e.g.\ one of the $(N-1)$ $U(1)$ factors in a generic point of the Coulomb branch. In 3D, the Hodge dual of the field strength tensor, $\star F$, is a one-form which may locally be written with respect to a scalar $\gamma$,
\begin{align}
\left[\star F\right]_i = \left[d\gamma\right]_i.
\label{eq:app:dual:fieldstrength}
\end{align}
The field $\gamma$ is known as the dual photon and encodes the degrees of freedom of $A_i$.
Due to charge quantization, $\gamma$ is periodic and thus takes values on $S^1$. From this it follows that the topology of the 3D $\mathcal N=2$ Coulomb branch for a $U(1)$ gauge theory is $\mathbb{R}\times S^1$. It is useful to combine the scalars into a complex modulus, 
\begin{align}
\phi = \sigma + i\gamma.
\label{eq:app:complex:coulomb}
\end{align}

\subsection{Linear superfields and duality}

Recall that the dual field strength $\mathcal J = \star F$ is a one-form (\ref{eq:app:dual:fieldstrength}). Maxwell's equations further imply that $\mathcal J$ is divergence free, $\partial^i \mathcal J_i =0$, so that 3D gauge theories carry a global topological symmetry, $U(1)_{\mathcal J}$, whose conserved current is exactly $\mathcal{J}_{i}$. 
In SUSY, conserved currents belong to a linear superfield, $\Sigma$, which satisfy $D^2\Sigma=\bar{D}^2\Sigma=0$. Because the $U(1)_{\mathcal{J}}$ is generated by the $\star F$, we may describe the vector superfield (\ref{app:SUSY:vec}) with an equivalent linear superfield,
\beq\label{app:SUSY:linear}
\Sigma\equiv-\frac{i}{2}\epsilon^{\alpha\beta}\bar{D}_\alpha D_\beta V=\sigma+\theta\bar{\lambda}+\bar{\theta}\lambda+\frac{1}{2}\theta\gamma^i\bar{\theta}\mathcal{J}_i+i\theta\bar{\theta}D+\frac{i}{2}\bar{\theta}^2\theta\gamma^i\partial_i\lambda-\frac{i}{2}\theta^2\bar{\theta}\gamma^i\partial_i\bar{\lambda}+\frac{1}{4}\theta^2\bar{\theta}^2\partial^2\sigma.
\eeq
In fact, the complex scalar (\ref{eq:app:complex:coulomb}) may be understood as the lowest component of a chiral superfield $\Phi$ that is dual to the linear superfield (\ref{app:SUSY:linear}). To see this, write the effective Lagrangian for a linear superfield is a real function
\beq
\mathcal{L}_\text{eff}=\int d^4\theta f(\Sigma).
\eeq
$\mathcal{L}_\text{eff}$ can further be written as a function of a general real superfield $\Sigma'$ with the addition of Lagrange multiplier chiral superfields $\Phi$ and $\Phi^\dag$
\beq
\mathcal{L}_\text{eff}=\int d^4\theta\left(f(\Sigma')-(\Phi+{\Phi^\dag})\frac{\Sigma'}{2\pi}\right).
\eeq
The equations of motion for $\Phi$ and $\Phi^\dag$ simplify enforce the linear superfield conditions on $\Sigma$, $D^2\Sigma=\bar{D}^2\Sigma=0$.
If one instead performs the path integral over $\Sigma'$, one obtains the condition
\begin{align}
\Phi + \Phi^\dag = 2\pi \frac{\partial f(\Sigma')}{\partial \Sigma'}
\end{align}
from which one may write
\beq
\mathcal{L}_\text{eff}=\int d^4\theta\; K(\Phi + \Phi^\dag),
\eeq
where $K$ is the Legendre transform of $f$. This is now a dual description of the vector superfield---encoded into a linear superfield---in terms of a chiral superfield whose lowest components are
\begin{align}
\left(\Phi+{\Phi^\dag}\right)|_{\theta=0} & =2{\rm Re}[ \phi]\equiv 2 \varphi
&
\left(\Phi-{\Phi^\dag}\right)|_{\theta=0} & =2i{\rm Im} [\phi]\equiv2i \gamma.
\end{align}
At tree level, $f(\Sigma)\sim \Sigma^2/g^2$ and the duality straightforwardly reproduces the results above,
\begin{align}
\varphi &\sim 2\pi\sigma/g^2
&
\partial_i\gamma & \sim\epsilon_{ijk}F^{jk}.
\end{align}
In other words, the scalar component of the vector superfield is identified with the real part of the complex scalar in $\Phi$, while its imaginary part can be identified with the dual photon.

The above discussion can naturally generalize to the ``bulk'' of the Coulomb branch of non-Abelian theories where $SU(N)\to U(1)^{N-1}$. There we can define $(N-1)$ topological $U(1)$s and $(N-1)$ dual photons. Therefore in the bulk of an $SU(N)$ Coulomb branch the vector superfield (\ref{app:SUSY:vec}) is dual to $(N-1)$ chiral superfields.


\section{3D Lagrangians}
\label{app:action}

In this section we summarize actions for matter and gauge superfields. 

\subsection{Gauge fields}
\label{app:action:gauge}

From a vector superfield $V$ (\ref{app:SUSY:vec}) one may use the gauge-invariant combination $W_\alpha=-\frac{1}{4}\bar{D}^2 e^{-V}D_\alpha e^V$ to construct the Yang-Mills action
\begin{align}\label{app:SUSY:vector:action}
S_\text{YM}&=\frac{1}{g^2}\int d^3x\, d^2\theta \; \left({\rm Tr}\, W_\alpha W^\alpha + \text{c.c.}\right)\\\nonumber
&=\frac{1}{g^2}\int d^3x\; {\rm Tr}\left(\frac{1}{4}F_{ij}F^{ij}+\mathscr{D}_i\sigma\mathscr{D}^i\sigma+D^2+\lambda^\dag\gamma^i\mathscr{D}_i\lambda\right).
\end{align}
As explained above, one may use an equivalent description in terms of the linear superfield $\Sigma$ (\ref{app:SUSY:linear}) for which the action is,
\beq\label{app:SUSY:linear:action}
S_\text{YM}=\frac{1}{g^2}\int d^3x\, d^4\theta\; \Sigma^2
\eeq
This is completely equivalent to (\ref{app:SUSY:vector:action}) once the $d^4\theta$ integral is performed.

In 3D, the Yang-Mills action is not the only gauge invariant combination of the gauge fields. Chern-Simons terms can also be added to the action. While we do not discuss theories with non-vanishing Chern-Simons terms in any detail, we remark that the supersymmetric generalization of these Chern-Simons terms is
\beq
S_\text{CS}=\int d^3x \; {\rm Tr}\left[\epsilon^{ijk}\big(A_i\partial_j A_k+i\frac{2}{3}A_iA_jA_k\big)+2D\sigma-\lambda^\dag\lambda\right],
\eeq 
which, in the Abelian case, can be written in a simple form involving both the vector and the linear multiplet:
\beq
S_\text{CS}\equiv\int d^3x \, d^4\theta \; \Sigma V.
\eeq
The existence of such terms in the theory drastically changes the physics. We refer to \cite{Intriligator:2013lca} for a recent detailed study of such theories. Fayet-Iliopoulos terms behave in the same way as they do in 4D and can be written for Abelian 3D theories.

\subsection{Matter fields}

The action for chiral superfields $Q$ is
\beq
S_\text{chiral}=\int d^3x \, d^4\theta \; K(Q,Q^\dag) +\int d^3x \, \left[ d^2\theta\, W(Q)+ \text{c.c.}\right]
\eeq
for K\"ahler potential $K(Q,Q^\dag)$ and superpotential $W(Q)$. In particular, for SUSY gauge theories the Kahler potential is $Qe^VQ^\dag$ so that the kinetic term is
\begin{align}\label{app:SUSY:matter}
\mathcal L_\text{kin.}
=
\left|\mathscr{D}_i\phi_Q\right|^2
+\phi_Q^\dag\sigma^2\phi_Q
+i\phi^\dag_Q D\phi_Q
+i\psi^\dag\gamma^i\mathscr{D}_i\psi
-i\psi^\dag\sigma\psi
+i\phi_Q^\dag\lambda^\dag\psi
-i\psi^\dag\lambda\phi_Q
+\left|F\right|^2,
\end{align}
where $\mathscr{D}_i$ is the Dirac operator and the subscript $\phi_Q$ indicates the lowest scalar component of the chiral superfield $Q$. We note that $\langle\sigma\rangle\neq0$ induces a supersymmetric mass term for $Q$. 

\subsection{Real and Complex Masses}

In 3D there are two different types of mass terms one may write for a chiral superfield $Q$. For vector-like theory we can write down a holomorphic mass adding a quadratic term to the super potential:
\beq
W_{m_{\mathbb{C}}}=m_{\mathbb{C}}Q\bar{Q}
\label{eq:complex:mass}
\eeq
because $m_{\mathbb{C}}$ is complex, this is known as a complex mass term and is the analog of the usual mass term in four dimensions.
Alternately, as noted in (\ref{app:SUSY:matter}), we may introduce a mass when $\langle \sigma \rangle \neq 0$. This is known as a {real mass} term and can be understood by modifying the K\"ahler potential,
\beq
\int d^3xd^4\theta Q e^{m_{\mathbb{R}}\theta^2} Q^\dag\sim \int d^3x\left( \frac{m_{\mathbb{R}}^2}{2}|\phi_Q|^2+im_{\mathbb{R}}\epsilon^{\alpha\beta}\bar{\psi}_\alpha\psi_\beta\right).
\eeq
Observe that the complex mass preserves parity while the real mass breaks parity, (\ref{eq:SUSY:3D:parity}).
The physical mass of the chiral superfield is $m=\sqrt{m_{\mathbb{R}}^2+m_{\mathbb{C}}^2}$.

A real mass can be induced by weakly gauging an exact global symmetry and fixing the weakly gauged vector superfield $\tilde{V}$ into a SUSY-preserving background configuration
\beq
\tilde{\sigma}=i\frac{m_{\mathbb{R}}}{g},\quad \tilde{A}_i=\tilde{\lambda}=\tilde{\bar{\lambda}}=\tilde{D}=0.
\eeq
Because the global symmetries of dual theories must coincide, one may perform a real mass deformation on one theory and straightforwardly map it to the dual theory.
This is used extensively in this paper and is explained thoroughly in Section \ref{check:duals}.

\section{Zero modes and Callias index theorem}
\label{app:callias}

Counting fermionic zero modes in an instanton/monopole background is a very useful tool to study the Coulomb branch of 3D $\mathcal{N}=2$ SUSY gauge theories. In this appendix we detail how to count these zero modes and derive (\ref{CoulombSU(4)}).

\subsection{Instanton/Monopole backgrounds}

The scalar component $\sigma$ of the vector superfield (\ref{app:SUSY:vec}) plays the role of and adjoint Higgs and admits nontrivial 3D instanton configurations. These 3D instantons can be understood as 4D monopoles.
For an $SU(2)$ gauge group in 4D with an adjoint Higgs $\sigma$, the 't Hooft-Polyakov monopole in singular gauge is
\begin{align}
\sigma=f(r,v)\tau^3,&\qquad f(r,v)=v \coth vr-\frac{1}{r}\\
A_i=\omega_i(r,v)\tau^3,&\qquad \omega_i(r,v)=\epsilon_{ij3}\hat{r}^j\left(\frac{1}{r}+\frac{v}{\sinh vr}\right)
\end{align}
where $\tau^3={\rm diag}(1,-1)$ is the third Pauli matrix to avoid ambiguity with the adjoint Higgs. The asymptotic behavior of the solution is
\beq
\sigma\big|_\infty=v\left(1-\frac{1}{vr}\right) \tau^3+\mathcal{O}\left(\frac{1}{r^2}\right),\qquad A_i\big|_{\infty}=\epsilon_{ij3}\frac{\hat{r}^j}{r}\tau^3+\mathcal{O}\left(\frac{1}{r^2}\right).
\eeq

The $SU(2)$ monopole solution generalizes straightforwardly to $SU(N)$ \cite{Weinberg:1980mo,Weinberg:1982mo}. There are $(N-1)$ independent monopoles which are labeled by integers $(n_1,n_2,...,n_{N-1})$, indicating the monopole charges. The asymptotic behavior is
\beq\label{app:callias:asympt}
\sigma\big|_\infty=\sigma_0-\frac{1}{r}\sum_{I=1}^{N-1}n_I\Big(\alpha_I\cdot H_I\Big)+\mathcal{O}\left(\frac{1}{r^2}\right),\qquad A_i\big|_{\infty}=\epsilon_{ij3}\frac{\hat{r}^j}{r}\sum_{I=1}^{N-1}n_I\Big(\alpha_I\cdot H_I\Big)+\mathcal{O}\left(\frac{1}{r^2}\right),
\eeq
where $\sigma_0\equiv{\rm diag}(\sigma_1,\sigma_2,...,\sigma_N)=\sum\sigma_0^IH_I$. Here $H_I$ are the generators of the Cartan sub-algebra, $\alpha_I$ are simple roots, and $\alpha_I\cdot H_I={\rm diag}(0,0...,1,-1,...,0)$. Further defining $g_0\equiv\sum n_I \big(\alpha_i\cdot H_I\big)={\rm diag}\big(n_1,n_2-n_1,...,-n_{N-1}\big)$, one may write (\ref{app:callias:asympt}) in a more concise form
\beq\label{app:callias:asympt2}
\sigma\big|_\infty=\sigma_0-\frac{g_0}{r}+\mathcal{O}\left(\frac{1}{r^2}\right),\qquad A_i\big|_{\infty}=\epsilon_{ij3}\frac{\hat{r}^j}{r}g_0+\mathcal{O}\left(\frac{1}{r^2}\right)
\eeq
upon rotating the time-coordinate in Euclidean spacetime onto the direction that is to be compactified and then dimensionally reducing to 3D, the solutions above become 3 dimensional instantons. We refer to these configurations as instanton/monopole solutions.

\subsection{Callias Index Theorem}

We may now state the Callias index theorem. For a fermion in representation $\mathcal R$ with weights $w_i$, the number of zero modes in an instanton/monopole background, (\ref{app:callias:asympt2}), is
\beq\label{app:callias:callias}
N=\frac{1}{2}\sum_{i}{\rm sign}(w_i\cdot \sigma_0)(w_i\cdot g_0).
\eeq 
For a careful proof see the original paper \cite{Callias:1978it} or \cite{deBoer:1997kr} for a more physical derivation. From the sign function one can see that conjugate representations have the same number of zero modes.


\subsection{Fundamental representation}

Let us computer the number zero modes of a fermion in the fundamental representation. We indicate the weights of the fundamental as $\{\nu_i\}$, where $i=1,...,N$. In this case we have
\begin{align}
\nu_i\cdot\sigma_0&=\sigma_i,
&
\nu_i\cdot g_0&=\big(g_0\big)_i=n_i-n_{i-1}\quad {\rm with}\quad n_0=n_N=0.
\end{align}
The number of zero modes is 
\beq
N_{\f}=\frac{1}{2}\sum_{i=1}^N{\rm sign}(\sigma_i)\big(n_i-n_{i-1}\big),
\eeq
which is the standard result that each fundamental fermion has $n_i$ zero modes in the region $\sigma_i>0>\sigma_{i+1}$ \cite{deBoer:1997kr,Aharony:1997bx,Poppitz:2008it}. In each region only the $i$-th fundamental instanton/monopole configuration contributes to the number of zero modes. Thus, in (\ref{CoulombSU(4)}), each $\f$ and $\af$ provide $n_1$ zero modes in region I, $n_2$ zero modes in region II and III, and $n_3$ in region IV.

\subsection{Adjoint representation}

Now consider the number of gaugino zero modes. 
The weights of the adjoint representation are the roots $\beta_{ij}\equiv \nu_i-\nu_j$, which we have expressed in terms of the fundamental weights, $\nu_i$.
The number of zero modes is 
\begin{align}
N_{\rm adj}&=\frac{1}{2}\sum_{i,j=1}^N{\rm sign}(\sigma_i-\sigma_j)\big[(n_i-n_{i-1})-(n_j-n_{j-1})\big]
\\
&=\sum_{i=1}^{N-1}\sum_{j=1}^{N-1}n_i\left[{\rm sign}(\sigma_i-\sigma_j)-{\rm sign}(\sigma_{i+1}-\sigma_{j+1})\right]
\\
&=\sum_{j=1}^{N-1}2 n_j.
\end{align}
Therefore the number of gaugino zero modes is independent of the values of the $\sigma_i$. This result matches with \cite{deBoer:1997kr,Aharony:1997bx,Poppitz:2008it}. 

\subsection{Antisymmetric representation}

For antisymmetric matter, the weights can be labelled with two indices $\left(w_{\rm anti}\right)_{ij}=\nu_i+\nu_j$ with $i\neq j$. The number of zero modes is thus
\begin{align}
N_{\rm anti}&=\frac{1}{2}\sum_{i\neq j}^N{\rm sign}(\sigma_i+\sigma_j)\big[(n_i-n_{i-1})+(n_j-n_{j-1})\big].
\end{align}
This depends on the sign of $\sigma_i+\sigma_j$. Referring to (\ref{CoulombSU(4)}), in regions I and II $(\sigma_1+\sigma_2;\sigma_1+\sigma_3;\sigma_1+\sigma_4)>0$ and $(\sigma_2+\sigma_3;\sigma_2+\sigma_4;\sigma_3+\sigma_4)<0$. Therefore each antisymmetric fermion has $N_{\rm anti}^\text{I, II}=2n_1$ zero modes. In regions III and IV both $\sigma_2+\sigma_3$ and $\sigma_1+\sigma_4$ flip sign. Inserting this into the above formula yields $N_{\rm anti}^\text{III, IV}=2n_3$. 
The zero mode counting for the fundamental and the adjoint representations matches the results in the literature. On the other hand, our results for antisymmetrics are obtained in a generic point on the Coulomb branch and are thus expected to differ from the conclusions of \cite{Shifman:2008conf} where the zero mode counting is performed at a center-symmetric point.

\section{Dimensional reduction of 4D dualities}
\label{app:dim:reduction}

We summarize the main features of the dimensional reduction of 4D dualities
\cite{Aharony:2013dha,Aharony:2013hda} focusing in particular on dimensional reduction of {s-confining theories}.
Consider a 4D duality between an `electric' theory $A_4$ and a `magnetic' theory $B_4$. One may then dimensionally reduce each theory in two steps:
\begin{enumerate}
\item Compactify a space-like dimension: $\mathbb{R}^4\to \mathbb{R}^3\times S^1$.
\item Reduce the size of the compact dimension to zero $\mathbb{R}^3\times S^1  \xrightarrow[r\to0]{}\mathbb{R}^3$
\end{enumerate}
This procedure indeed reduces the 4D theories into 3D theories. However the 3D theories  so obtained, $(A_3, B_3)$, are not duals of one another. This can be understood intuitively when $A_4$ and $B_4$ are Seiberg duals. These are related by an infrared duality that is valid at energies much lower than the confinement scales of the dual theories\footnote{We implicitly assume that both theories are asymptotically free. This assumption plays no role in what follows and the conclusions apply also to the case where the magnetic theory is IR free.}, 
\begin{align}
E\ll\Lambda_{A_4}, \Lambda_{B_4}
&&
\Lambda=\mu \, e^{-\frac{8\pi^2}{g^2(\mu)b}}.
\end{align}
%
%
%
%
After compactification, the effective 3D gauge coupling depends on the size of the compact dimension,
$g^2_4=2\pi r g^2_3$.
The limit $r\to 0$ thus corresponds to $\Lambda_{A_4}, \Lambda_{B_4} \to 0$. This means that the IR limit $E \ll \Lambda_{A_4,B_4}$ is not valid in the na\"ive compactification and the duality does not hold.

In order to find a 3D version of 4D Seiberg duality, one must examine the compactified theory at finite radius,
\begin{align}
E \ll \Lambda_{A_4}, \Lambda_{B_4} \ll \frac{1}{r}.
\end{align}
The theories on $\mathbb{R}^3\times S^1$ are intrinsically different from the na\"ive dimensional reduction because the compact $S^1$ direction allows an extra topological configuration 
that is distinct from the purely 3D topology described in Section~\ref{sec:CB} and Appendix~\ref{app:callias}. 
This additional configuration is a a Kaluza-Klein (KK) monopole that wraps around the circle with a twist \cite{Lee:1997mo,Lee:1998mo,Lee:1998ca}. Like the other instanton/monopole configurations, the KK monopole also generates a non-perturbative contribution to the superpotential \cite{Affleck:1982as},
\begin{align}
\label{eq:instel} 
W_{S^1}=\eta Y,
&&
\eta=\Lambda^b,
&&
Y=\exp\left(\frac{\sigma_1-\sigma_N}{g_3^2}+i(a_1-a_N))\right),
\end{align}
where $Y$ is the chiral operator introduced in (\ref{eq:Y:prod:Yi}). The superpotential $W_{S^1}$ must be included in the compactified theory to preserve the duality. In the $r\to 0$ limit,  $\eta\to 0$ and this term in the superpotential disappears \cite{Davies:2000mo}. 
In SQCD with $F>N$, the resulting IR duality between intermediate compactified theories is
\begin{align}\label{eq:dual}
\begin{array}{c}
SU(N)\ \text{ with } F(\f+\af)\\
W=\eta Y
\end{array}\quad\longleftrightarrow\quad
\begin{array}{c}
SU(F-N)\ {\rm with}\ F(\f+\af)\\
W=qM\bar{q}+\tilde{\eta}\tilde{Y}
\end{array},
\end{align}
with $\eta\hskip1pt\tilde{\eta}=(-1)^{F-N}$.
%
%
%
The superpotential term from the KK monopole configuration can also be decoupled. This can be done by decoupling one flavor in the compactified theory with a real mass deformation \cite{Aharony:2013dha}.

\begin{figure}
\begin{center}
\includegraphics[width=.87 \textwidth]{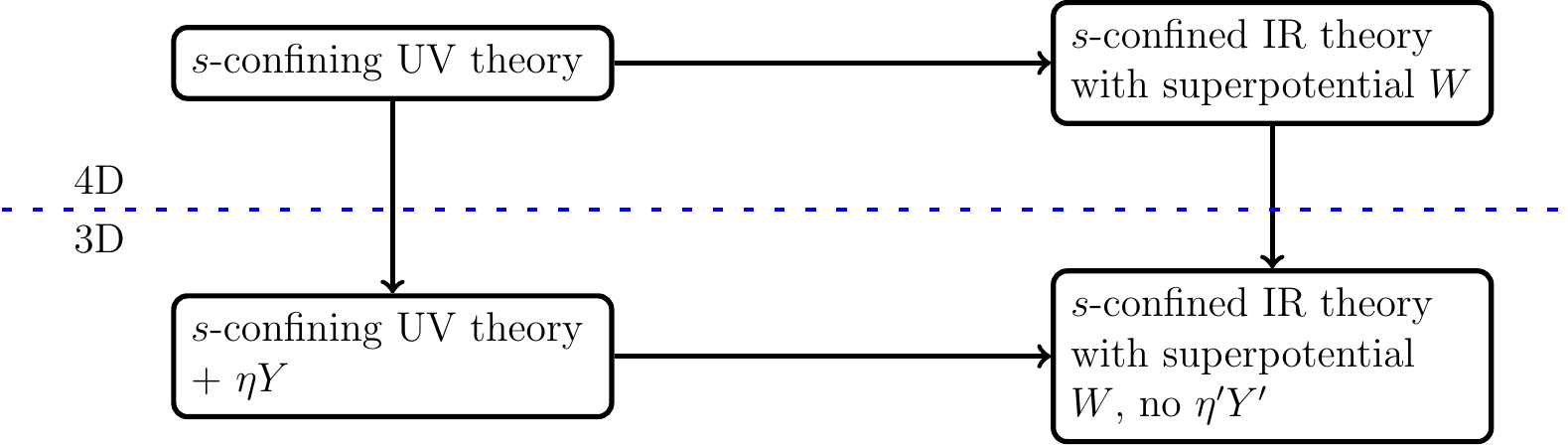}
\end{center}
\caption{If we reduce {s-confining} theories, there is no $\tilde{\eta}\tilde{Y}$ in the magnetic side.}
\label{fig:sconfinement:4d:3d}
\end{figure}

S-confining theories are a particular set of dual 4D supersymmetric theories where the electric theory, $A_4$, confines and the magnetic degrees of freedom describe the resulting composite degrees of freedom. The dual theory, $B_4$< has no gauge group. These have the following properties \cite{Csaki:1996sm, Csaki:1998th}:
\begin{enumerate}
\item The IR physics is described exactly by gauge invariant composites.
\item A confining superpotential is dynamically generated.
\item The origin of the classical moduli space, where all the global symmetries of the electric theory are unbroken, is also an IR vacuum in the quantum theory.
\end{enumerate}
%
%
%
%
For these theories, the dimensional reduction algorithm follows as it did above with the $\mathbb{R}^3\times S^1$ magnetic theory having no $\tilde\eta \tilde Y$ term as the magnetic theory has no gauge group, as shown in Fig.~\ref{fig:sconfinement:4d:3d}.
%
One may perform consistency checks for the resulting 3D duality.
In what follows, we set $\Lambda=1$ for notational simplicity.

%
%
%
%
%
%

\subsection{Check 1}

From the procedure above, SQCD with $F=N+2$ flavors on $\mathbb{R}^3\times S^1$ is dual to $SU(2)$ with $F=N+2$ flavors and a superpotential
\begin{align}
W = \bar q M q + \tilde\eta\tilde Y.
\label{eq:app:reduction:check1:a}
\end{align} 
Deforming the electric theory by a complex mass (\ref{eq:complex:mass}) for one flavor then leads to SQCD with $F=N+1$ and $W=(\eta/m)Y|_{F=N+1}\equiv \eta'Y'$. From Fig.~\ref{fig:sconfinement:4d:3d} we expect that this theory is described in the IR by the 4D s-confining superpotential with no extra $\tilde{\eta}\tilde{Y}$ term:
\begin{align}\label{Check1}
SU(N)\text{ with } W= \eta' Y'\quad\&\quad F=N+1
&
\qquad\longleftrightarrow\qquad
W = B M\bar B - \det M.
\end{align}
To see how this correspondence arises in a 3D description, we note that a holomorphic mass for the $(N+2)^\text{th}$ electric quark flavor maps onto a tadpole for the $(N+2)^\text{th}$ diagonal element of the magnetic meson. As a result the last flavor of dual quark acquire a VEV $q_{N+2}\bar q_{N+2}=-m$ completely breaking magnetic $SU(2)$ group.
After identifying the remaining dual quarks with the baryons of $N+1$ flavor theory, the low energy superpotential becomes\footnote{Unless explicitly noted, from now on $M$ refers only to meson composites of light electric quarks.}
\begin{equation}\label{Check1low}
 W=BM\bar B+\tilde \eta\tilde Y\,.
\end{equation}
An interpretation of $\tilde \eta \tilde Y$ term generated by the KK instanton is somewhat non-trivial because $SU(2)$ is completely broken.
Comparing to 4D physics, we know that this term is proportional to $\det M$ since it must reproduce the effect of 4D instanton. It is, however, instructive to obtain this result from a purely 3D perspective. To this end, consider a limit of small $m$ and large $M$, such that\footnote{Since classical constraints are not modified quantum mechanically in this theory, one must have $\mathrm{rank}(M)\le N$ in the vacuum. Thus, a superpotential for $M$ must be generated dynamically.} $\mathrm{rank }M=N+1$.
In this regime, the low energy physics is described by a single flavor $SU(2)$. It is known that such a theory has a quantum modified constraint $\tilde Y_\mathrm{low}=1/(q_{N+2}\bar q_{N+2})$ \cite{Aharony:1997bx}. Matching quantum KK instanton-monopole operators of the high and low energy theories, we find
\begin{equation}
 \tilde Y=\det M \; \tilde Y_\mathrm{low}\,.
\end{equation}
Finally, holomorphy guarantees that this result is also valid in the large $m$ regime where $q_{N+2}\bar q_{N+2}=-m$. Thus, after appropriate rescalings, the superpotential (\ref{Check1low}) becomes
\begin{equation}
 W=BM\bar B-\mathrm{det} M\,,
\end{equation}
in full agreement with (\ref{Check1}).

\subsection{Check 2}

A second non-trivial check is to start instead with $SU(N)$ with $(N+1)$ flavors and $W=\eta Y$. Because of the lack of the KK monopole, such theory should be dual to an \emph s-confined theory with super-potential 
\begin{align}
W_{\rm mag}=B^iM_i^j\bar{B}_j-\det M.
\end{align}
Adding a real mass to $\mathcal{Q}$ and $\bar{\mathcal{Q}}$ and taking $m_{\mathbb{R}}\to\infty$ decouples the instanton term in the electric theory (for more details see the appendix or \cite{Aharony:2013dha} ). In order not to induce extra Chern-Simons term, we need to assign $m_{\mathbb{R}}^{\mathcal{Q}}=-m_{\mathbb{R}}^{\bar{\mathcal{Q}}}$. This is easily done by weakly gauging $\big[SU(N+1)_L\times SU(N+1)_R\big]_D\times U(1)_B$. The masses induced in the magnetic theory are:
\begin{align}
m_B = 
\begin{pmatrix}
m_\mathbb{R}\\
0\\
\vdots\\
0
\end{pmatrix}
&&
m_{\bar B} = 
\begin{pmatrix}
-m_\mathbb{R}\\
0\\
\vdots\\
0
\end{pmatrix}
&&
m_M
= 
\begin{pmatrix}
0 & \neq 0\\
\neq 0 & 0
\end{pmatrix}
\end{align}
Identifying the $(N+1)\times (N+1)$-th component of the meson field with the Coulomb branch operator of the electric theory, and decoupling the massive fields, we obtain
\begin{align}
W_{\rm mag}=Y(B\bar{B}-\det M)
\end{align}
which matches the known IR description of the $SU(N)$ with $N$ flavors. Therefore again the initial assumption of no $\tilde{\eta}\tilde{Y}$ in the magnetic theory is consistent from a fully 3D perspective.



\small
\bibliographystyle{utphys}    
\bibliography{sconfine3D}     

\end{document}